\RequirePackage{fix-cm}
\documentclass[smallextended]{svjour3}       
\usepackage{graphicx}
\usepackage{epsfig,amsmath,amssymb,mathtools,amscd}
\usepackage{mathptmx} 
\usepackage{enumerate}
\usepackage{csquotes}
\usepackage{bm}
\usepackage[active]{srcltx}
\def\fext{pdf}    

\def\eg{{e.~g.} }\def\ie{{i.~e.} }
\mathchardef\minus="002D

\def\<{\langle}\def\>{\rangle}

\def\v#1{\boldsymbol{\mathrm #1}} 
\def\Z{\mathbb Z} 

\def\Reals{\mathbb{R}}\def\Cmplx{\mathbb{C}}

\def\transf#1{\mathcal{#1}}
\def\tA{\transf A}

\def\L2{{\mathcal L}_2}

\def\d#1 {\mathop{\!\! \mathrm{d}#1}\,}
\def\df#1#2 {\!\!\frac{\mathop{\mathrm{d}#1}}{#2}\,}
\def\SU{\mathbb{SU}}

\def\bvec#1{\mathbf{#1}}
\def\v#1{\bvec{#1}}
\def\bk{\bvec k}
\def\bh{\bvec h}

\def\bn{\bvec n}

\def\bg{\bvec g}

\def\bn{\bvec n}

\def\bbeta{{\bvec{\beta}}}
\def\Brill{\mathsf B}

\def\d{\operatorname{d}}
\def\Exp{\operatorname{Exp}}

\def\spc#1{{\cal{#1}}}
\def\sH{\spc H}
\def\sK{\spc K}
\def\Span{\mathsf{Span}}
\journalname{Int J Theor Phys}
\begin{document}

\title{Physics without physics\thanks{Work supported by the Templeton
    Foundation under the project ID\# 43796 {\em A Quantum-Digital Universe}.}}
\subtitle{The power of information-theoretical principles\\
\flushright{\rm\em a tribute to David Finkelstein}}
\titlerunning{Physics without physics and the power of information-theoretical principles}
\author{Giacomo Mauro D'Ariano}
\institute{Giacomo Mauro D'Ariano \at
              University of Pavia, {\em QUit Group}, Dipartimento di Fisica and INFN gruppo IV, via
              Bassi 6, 27100 Pavia \\ 
              Tel.: +39 0382 987484, \email{dariano@unipv.it}\\
\\             \emph{Temporary address:}\\ 
Northwestern University, Department of Electrical and Computer Engineering\\
Tech. Institute, 2145 Sheridan Road, Evanston, IL 60208 USA}
\date{Received: date / Accepted: date}
\maketitle

\begin{abstract}
  David Finkelstein was very fond of the new information-theoretic paradigm of physics advocated by
  John Archibald Wheeler and Richard Feynman.  Only recently, however, the paradigm has concretely
  shown its full power, with the derivation of quantum theory \cite{chiribella2011informational,QTfromprinciples} and
  of free quantum field theory \cite{PhysRevA.90.062106,bisio2013dirac,Bisio2015244,Bisio2016} from
  informational principles. The paradigm has opened for the first time the possibility of avoiding
  physical primitives in the axioms of the physical theory, allowing a re-foundation of the whole
  physics over logically solid grounds. In addition to such methodological value, the new
  information-theoretic derivation of quantum field theory is particularly interesting for
  establishing a theoretical framework for quantum gravity, with the idea of obtaining gravity
  itself as emergent from the quantum information processing, as also suggested by the role played by information in
  the holographic principle \cite{susskind1995world,bousso2002holographic}.

  In this paper I review how free quantum field theory is derived without using mechanical primitives,
  including space-time, special relativity, Hamiltonians, and quantization rules. The theory is simply provided by the
  simplest quantum algorithm encompassing a countable set of quantum systems whose network of
  interactions satisfies the three following simple principles: homogeneity, locality, and isotropy.

  The inherent discrete nature of the informational derivation leads to an extension of quantum
  field theory in terms of a quantum cellular automata and quantum walks. A simple heuristic argument  
  sets the scale to the Planck one, and the currently observed regime where discreteness is not visible 
  is the so-called ``relativistic regime'' of small wavevectors, which holds for all energies ever tested
  (and even much larger), where the usual free quantum field theory is perfectly recovered. 
  
  In the present quantum discrete theory Einstein relativity principle can be restated without 
  using space-time in terms of invariance of the eigenvalue equation of the automaton/walk under
  change of representations. Distortions of the Poincar\'e group emerge at the Planck scale, 
  whereas  special relativity is perfectly recovered in the relativistic regime. Discreteness, on 
  the other hand, has some plus compared to the continuum theory: 1) it contains it as a 
  special regime; 2) it leads to some additional features with GR flavor: the existence 
  of an upper bound for the particle mass (with physical interpretation as the Planck mass), and 
  a global De Sitter invariance; 3) it provides its own physical standards for 
  space, time, and mass within a purely mathematical adimensional context.
  
  The paper ends with the future perspectives of this project, and with an appendix containing biographic 
  notes about my friendship with David Finkelstein, to whom this paper is dedicated. 

\end{abstract}

\bigskip
\begin{flushright} 
{\em Beware the Lorelei of Mathematics.  Her song is beautiful.}

David Finkelstein
\end{flushright} 

\section{Introduction}
\label{intro}
The logical clash between General Relativity (GR) and Quantum Field Theory (QFT) is the
main open problem in physics. The two theories represent our best theoretical frameworks, and work
astonishingly well within the physical domain for which they have been designed. However, their
logical clash requires us to admit that they cannot be both correct. One could argue that there
must exist a common theoretical substratum from which both theories emerge as approximate effective
theories in their pertaining domains--though we know very little about GR in the domain of particle
physics.

What we should keep and what we should reject of the two theories? Our experience has thought us
that of QFT we should definitely keep the Quantum Theory (QT) of abstract systems, namely
the theory of the von Neumann book \cite{john1955mathematical} stripped of its ``mechanical'' part,
\ie the Schr\"odinger equation and the quantization rules. This leaves us with the description of
generic systems in terms of Hilbert spaces, unitary transformations, and observables.
In other words, this is what nowadays is also called Quantum Information, a research field indeed  very
interdisciplinary  in physics.

There are two main reasons for keeping QT as valid.  First, it has been never falsified in any
experiment in the whole physical domain--independently of the scale and the kind of system. This
has lead the vast majority of physicists to believe that everything must behave according to QT.
The second and more relevant reason is that QT, differently from any other chapter of physics, is
well axiomatized, with purely mathematical axioms containing no physical primitive. So, in a sense,
QT is as valid as a piece of pure mathematics. This must be contrasted with the mechanical part of the
theory, with the bad axiomatic of the so-called ``quantization rules'', which are extrapolated
and generalized starting from the heuristic argument of the Ehrenfest theorem, which in turn is
based on the superseded theory of classical mechanics, and with the additional problem of the ordering 
of canonical noncommuting observables.\footnote{The problem of ordering is avoided miraculously 
thanks to the fortuitous non occurrence in nature of Hamiltonians with products of conjugated observables.} 
No wonder then that the quantization procedure doesn't work well for gravity!

To what we said above we should add that today we know that the QT of von Neumann can be derived
from six information-theoretical principles \cite{chiribella2011informational,QTfromprinciples}, 
whose epistemological value is not easy to give up.\footnote{For short reviews, see also Refs. 
\cite{Giuliobook,shortreviewQT}.} On the contrary, it is the mechanical part of
QFT that rises the main inconsistencies, e~.g. the Malament theorem \cite{halvorson2002no}, which
makes any reasonable notion of particle untenable \cite{kuhlmann2015real}.

The logical conclusion is that what we need is a field theory that is {\em
  quantum ab initio}. But how to avoid quantization rules? The idea is simply to consider a countable
set of quantum systems in interaction, and to make the easiest assumptions on the topology
of their interactions. These are: locality, homogeneity, and isotropy. Notice that  we are not using
any mechanics, nor relativity, and not even space and time. And what we get?  We get: Weyl, Dirac
\cite{PhysRevA.90.062106}, and Maxwell \cite{Bisio2016}. Namely: we get free quantum field theory!

The new general methodology suggested to the above experience is then the following: 
1) no physical primitives in the axioms; 2) physics only as interpretation of the mathematics 
(based on experience, previous theories, and heuristics). In this way the logical coherence 
of the theory is mathematically guaranteed. In this review we will see how the 
proposed methodology can be actually carried out, and how  the informational paradigm has the potential of solving 
the conflict between QFT and GR in the case of special relativity, with the latter
emergent merely from quantum systems in interaction: Fermionic quantum bits at the
very tiny Planck scale. In synthesis the program is an {\em algorithmization of theoretical physics}, 
aimed to derive  the whole physics from quantum algorithms with finite complexity, 
upon connecting the algebraic properties of the algorithm with the dynamical features of the physical theory, 
preparing a logically coherent framework for a theory of quantum gravity.

\medskip

Section \ref{s:assumptions} is devoted to the derivation from principles of the quantum-walk theory.
More precisely, from the requirements of homogeneity and locality of the interactions of countably many
quantum systems one gets a theory of quantum cellular automata on the Cayley graph of a group $G$.
Then, upon restricting to the simple case of evolution linear in the discrete fields, the quantum automaton 
becomes what is called in the literature {\em quantum walk}. We further restrict to the case with physical 
interpretation in an Euclidean space, resorting to considering only Abelian $G$. 

In Section \ref{s:Abelian} the quantum walks with minimal field dimension that follow
from the principles of Sect. \ref{s:assumptions} are reported. These represent the Planck-scale version of
the Weyl, Dirac, and Maxwell quantum field dynamics, which are recovered in the relativistic
regime of small wavevectors. Indeed, the quantum-walk theory, being purely mathematical--and so 
adimensional--nevertheless contains its own physical LTM standards written in the intrinsic discreteness
and non-linearities of the theory. A simple heuristic argument based on the notion of mini black-hole 
(from a matching of GR-QFT) leads to the Planck scale. It follows that the relativistic regime
contains the whole physics observed up to now, including the most energetic events from cosmic rays.  

In addition to the exact dynamics in terms of quantum walks, a simple analytical method is also
available in terms of a dispersive Schr\"odinger equation, suitable to the Planck-scale physics for
narrow-band wave-packets. As a result of the unitarity constraint for the evolution, the particle mass 
turns out to be upper bounded (by the Planck mass), and has domain in a circle, corresponding to 
having also the proper-time (which is conjugated to the mass) as discrete. Effects due to discreteness 
that are in principle  visible are also analyzed, in particular a dispersive behavior of the vacuum, 
that can be detected by deep-space ultra-high energy cosmic rays.

Section \ref{s:SR} is devoted to how special
relativity is recovered from the quantum-walk discrete theory, without using space-time and
kinematics. It is shown that the transformation group is a non-linear
version of the Poincar\'e  group, which recovers the usual linear group in the relativistic limit of
small wavevectors. For nonvanishing masses generally also the mass gets involved in the
transformations, and the De Sitter group $SO(1,4)$ is obtained. 

The paper ends with a brief section on the future perspectives of the theory, and with an Appendix
about my first encounter with David Finkelstein.

Most of results reported in the present review have been originally published in Refs. 
\cite{PhysRevA.90.062106,bisio2013dirac,Bisio2015244,Bisio2016,reviewderiv,reviewanaly,mauro2012quantum,%
lrntz3d,FOPDP16,d2015virtually} coauthored with members of the QUit group in Pavia.

\section{Derivation from principles of the quantum-walk theory}\label{s:assumptions}
\medskip
\begin{flushright} 
{\em If you are receptive and humble, mathematics will lead you by the hand.}

Paul Dirac
\end{flushright} 
\medskip
The derivation from principles of quantum field theory starts from considering the unitary evolution
$\tA$ of a countable set $G$ of quantum systems, with the requirements of homogeneity, locality, 
and isotropy of their mutual interactions. These will be precisely defined and analyzed in following
dedicated subsections. All the three requirements are dictated from the general principle of minimizing
the algorithmic complexity of the physical law. The physical law itself is described by a finite 
quantum algorithm, and homogeneity and isotropy assess the universality of the law. 

The quantum system labeled by $g\in G$ can be either associated to an Hilbert space $\sK_g$, or to a
set of generators of a C${}^*$-algebra\footnote{The two associations can be connected through
  the GNS construction.}
\begin{equation}
\psi_g\equiv\{\psi_g^\nu\},\qquad g\in G,\quad \nu\in[s_g]:=\{1,2,\ldots, s_g\},\; s_g<\infty.
\end{equation}
The evolution occurs in discrete identical steps\footnote{More generally the map $\tA$ is an automorphism of the algebra.}
\begin{equation}\label{e:Q}
\tA\psi_g=U\psi_gU^\dag,\quad U\text{unitary,}
\end{equation}
describing the interactions among systems. When the unitary evolution is also {\em local}, namely
$\tA\psi_g$ is spanned by a finite subset $S_g\subset G$, then $\tA$ is called {\em Quantum Cellular
  Automaton}. We restrict to evolution linear in the generators, namely
\begin{align}\label{eq:Alinear}
\tA\psi_g=U\psi_gU^\dag=\sum_{g'}A_{g,g'}\psi_{g'},
\end{align}
where $A_{g,g'}$ is an $s_g\times s_{g'}$ complex matrix called {\em transition matrix}.  Here in
all respects the quantum cellular automaton is described by a unitary evolution on a (generally infinite) Hilbert space
$\sH=\bigoplus_{g\in G}\sH_g$, with $\sH_g=\Span\{\psi_g^\nu\}_{\nu\in[s_g]}$. In this case the
quantum cellular automaton is called {\em Quantum Walk}. Here the system simply corresponds to a
finite-dimensional block component of the Hilbert space, regardless the Bosonic/Fermionic nature of
the field. In the derivation of free quantum field theory from principles, the quantum walk
corresponds to the evolution on the single-particle sector of the Fock space, whereas for the
interacting theory a generally nonlinear quantum cellular automaton is needed. 
Simple generalization to Fock-space sectors with fixed number of particles are also possible.

\subsection{The quantum system: qubit, Fermion or Boson?}\label{s:bosonferm}
At the level of quantum walks, corresponding to the Fock space description of cellular quantum
automata (leading to free QFT in the nonrelativistic limit), it does not make any difference which
kind of quantum system is evolving. Indeed one can symmetrize or anti-symmetrize products of
wavefunctions, as it is done in usual quantum mechanics, or else just take products with no
symmetrization. Things become different when the vacuum is considered, and particles are
created and annihilated by operating with algebra generators on the vacuum state, as in the interacting theory. 
Therefore, as far as we are concerned with free QFT, which kind of quantum system should be used is 
a problem that can be safely postponed.

However, there are still motivations for adopting a kind of quantum system instead of another.
For example, a reason for discarding qubits as algebra generators is that there is no easy way of
expressing the operator $U$ making the evolution in Eq. (\ref{eq:Alinear}) linear,
whereas, when $\psi_g$ is Bosonic or Fermionic this is always possible choosing $U$
exponential of bilinear forms in the fields. On the other hand, a reason to chose Fermions instead
of Bosons is the requirement that the amount of information in a finite number of cells be finite,
namely one has finite information density in space.\footnote{Richard Feynman is reported to
  like the idea of finite information density, because he felt that: {\em ``There might be something
    wrong with the old concept of continuous functions.  How could there possibly be an infinite
    amount of information in any finite volume?''}  \cite{hey1998feynman}.}  The relation between
Fermionic modes and finite-dimensional quantum systems, say \emph{qubits} has been studied in the
literature, and the two theories have been proven to be computationally equivalent
\cite{Bravyi2002210}.  However, the quantum theory of qubits and the quantum theory of Fermions
differ in the notion of what are local transformations \cite{IJMP14,EPL14}, with local Fermionic
operations mapped into nonlocal qubit transformations and vice versa.

In conclusion, the derivation from informational principles of the fundamental particle statistics still
remains an open problem. One could promote the finite information density to the level of a
principle, or motivate the Fermionic statistics from other principles of the same nature of those in
Ref.  \cite{chiribella2011informational} (see \eg Refs. \cite{IJMP14,EPL14}), or derive the
Fermionic statistics from properties of the vacuum (\eg having a localized non-entangled vacuum in
order to avoid the problem of particle localization), and then recover the \emph{Bosonic} statistics
as a very good approximation, with the Bosonic mode corresponding to a special entangled state of
pairs of Fermionic modes \cite{Bisio2016}, as it will be reviewed in Subsect.  \ref{s:maxwell}. This
hierarchical construction will also guarantee the validity of the spin-statistic connection in QFT.

\subsection{Quantum Walks on Cayley graphs\protect\footnote{This
   subsection is based on results of Refs. \cite{PhysRevA.90.062106} and \cite{lrntz3d}.}}
\begin{figure}
\begin{center}
\includegraphics[width=.8\textwidth]{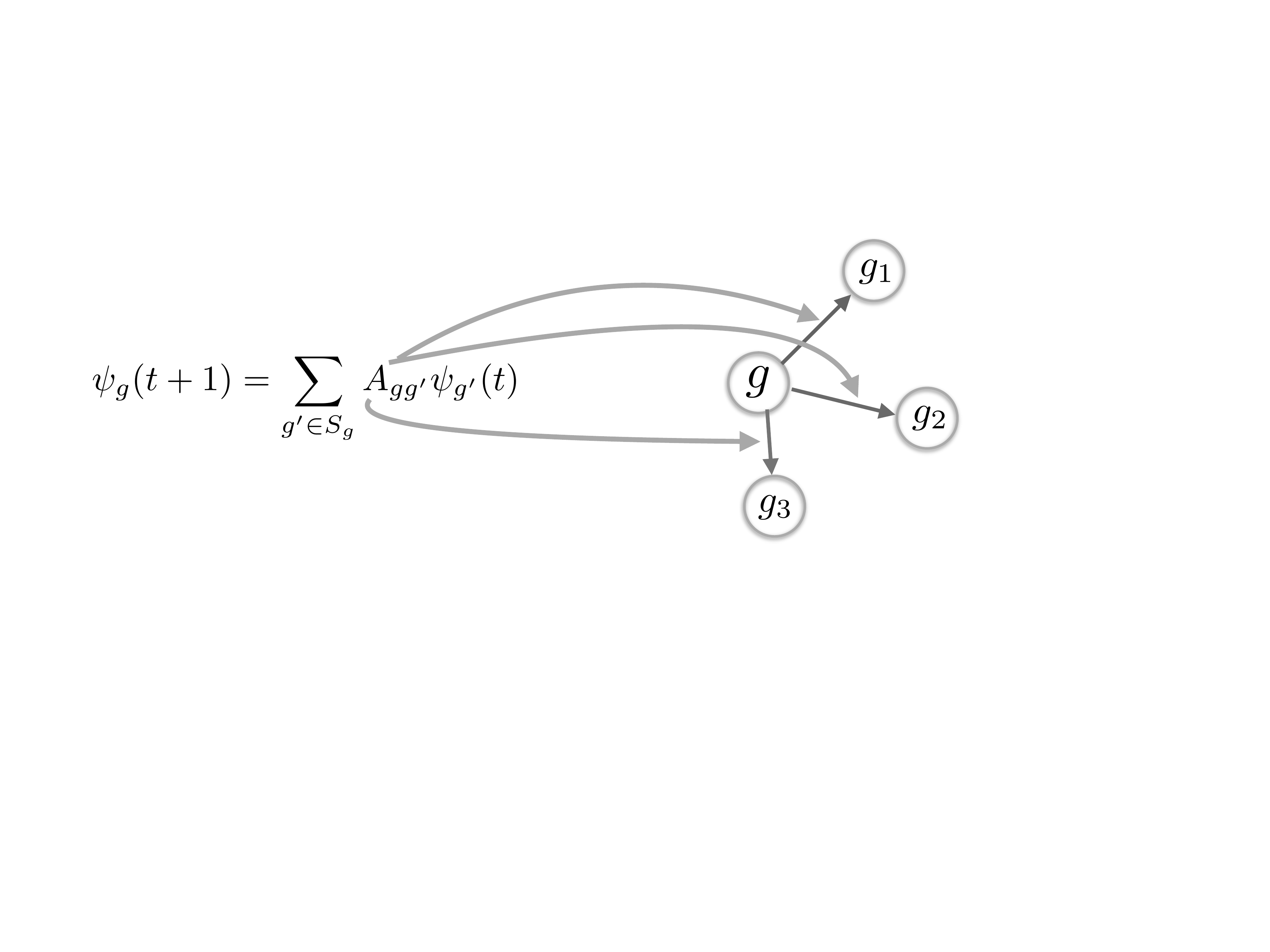}
\vskip -2.5cm
\caption{The linear Eq. (\ref{eq:Alinear}) endows the set $G$ with a {\em directed graph structure}.
  We build a directed graph with an arrow from $g$ to $g'$ wherever the two are connected by a
  nonnull matrix $A_{gg'}$ in Eq. (\ref{eq:Alinear}).\label{fig:arrowA}}
\end{center}
\end{figure}

The linear Eq. (\ref{eq:Alinear}) endows the set $G$ with a {\em directed graph structure}
$\Gamma(G,E)$, with vertex set $G$ and edge set $E=\{(g,g')|A_{g,g'}\neq0\}$ directed from $g$ to
$g'$ (see Fig. \ref{fig:arrowA}). In the following we will denote by $S_g:=\{A_{g,g'}\neq0\}$ the set
of non-null transition matrices with first index $g$, and by $N_g:=\{g'\in G|A_{g,g'}\neq 0\}$ the
{\em neighborhood} of $g$. 

\subsubsection{The homogeneity principle}\label{s:homo}
The assumption of homogeneity is the requirement that every two vertices are indistinguishable,
namely for every $g,g'\in G$ there exists a permutation $\pi$ of $G$ such that $\pi(g)=g'$ which
commute with any discrimination procedure consisting of a preparation of local modes followed by a
general joint measurement. In Ref. \cite{FOPDP16} it is shown that this is equivalent to the
following set of conditions 

\medskip
$\forall g\in G$ one has:
\begin{enumerate}[H1]
\item $s_g=s$;\label{i:s}
\item there exists a bijection $N_g\leftrightarrow N$ with a fixed set $N$;\label{i:N}
\item $S_g$ contains the same $s\times s$ transition matrices, namely
  $S_g=S:=\{A_{h_1}\}_{i=1}^{|N|}$;\label{i:color}  
\item $A_{g,g'}=A_{h_i}\in S$ $\Rightarrow$ $A_{g',g}=A_{h_j}\in S$;\label{i:invert}
\end{enumerate}

Condition H\ref{i:N} states that $\Gamma(G,E)$ is a \emph{regular graph}---\ie each vertex has the
same degree. Condition H\ref{i:color} makes $\Gamma(G,E)$ a {\em colored directed graph}, with the
arrow directed from $g$ to $g'$ for $A_{g,g'}=A_h\in S$ and the color associated to $h$.\footnote{If
  two transition matrices $A_{h_1}=A_{h_2}$ are equal, we conventionally associate them with two
  different labels $h_1\neq h_2$ in such a way that $\sum_{f\in
    N_{\pi(g)}}A_{\pi(g)f}\psi_{\pi^{-1}(f)}=\sum_{f\in N_{g}}A_{gf}\psi_{f}$. If such choice is not
  unique, we will pick an arbitrary one, since the homogeneity requirement implies that there exists
  a choice of labeling for which all the construction that will follow is consistent.}  Condition
H\ref{i:color} introduces the following formal action of symbols $h_i\in S$ on the elements $g\in G$
as
\begin{equation}
A_{gg'}=A_{h_i}\Rightarrow gh_i=g'.
\end{equation}
Clearly such action is closed for composition. From condition H\ref{i:invert} one has that
\begin{equation}
A_{g'g}=A_{h_j}\Rightarrow g'h_j=g,
\end{equation}
and composing the two actions we see that $gh_ih_j=g$, and we can write the label $h_j$ as
$h_j=:h_i^{-1}$. We thus can build the free group $F$ of words made with the alphabet $S$. Each word
corresponds to a path over $\Gamma(G,E)$, and the words $w\in F$ such that $gw=g$ correspond to
closed paths (also called {\em loops}). Notice that by construction, one has
$A_{\pi(g)\pi(f)}=A_{gf}=A_{h_i}$, which implies that $\pi(g)h_i=\pi(f)=\pi(gh_i)$, from which one
can prove that $f'w=\pi(f)w=\pi(fw)=\pi(f)=f'$ (see \cite{FOPDP16}). Thus
we have the following
\begin{enumerate}[H5]
\item\label{i:path} If a path $w\in F$ is closed starting from $f\in G$, then it is closed also starting from any
  other $g\in G$. 
\end{enumerate}
The subset $R\subset F$ of words $w$ such that $gw=g$ is obviously a group. Moreover $R$ is a normal
subgroup of $G$, since $gwrw^{-1}=(gw)rw^{-1}=(gw)w^{-1}=g$, namely $wrw^{-1}\in R$ $\forall w\in
F,\forall r\in R$. Obviously the equivalence classes are just elements of $G$, which means that
$G=F/R$ is a group. Pick up any element of $G$ as the identity $e\in G$.  It is clear that the
elements of the quotient group $F/R$ are in one-to-one correspondence with the elements of $G$,
since for every $g\in G$ there is only one class in $F/R$ whose elements lead from $e$ to $g$ (write
$g=ew$ for every $w\in F$, $w$ representing a path leading from $e$ to $g$). The graph $\Gamma(G,E)$
is thus what is called in the literature the {\em Cayley graph of the group} $G$ (see the 
definition in the following). The Cayley graph is in correspondence with a {\em presentation} of the group $G$. This
is usually given by arbitrarily dividing the set as $S=S_+\cup S_-$ with
$S_-:=S_+^{-1}$,\footnote{The above arbitrariness is inherent the very notion of group presentation
  and corresponding Cayley graph, and will be exploited in the following, in particular in the
  definition of isotropy.} and by considering a set $W$ of generators for the free group of loops
$R$. The group $G$ is then given with the presentation $G= \<S_+|W\>$, in terms of the set of its
{\em generators} $S_+$ (which along with their inverses $S_-$ generate the group by composition),
and in terms of the set of its {\em relators} $W$ containing group words that are equal to the
identity, with the goal of using these words in $W$ to establish if any two words of elements of $G$
correspond to te same group element. The relators can also be regarded as a set of generators for
$R$.  

The definition of \emph{Cayley graph} is then the following.

\paragraph{Cayley graph of $G$.} Given a group $G$ and a set $S_+$ of generators of the group, the
Cayley graph $\Gamma(G,S_+)$ is defined as the colored directed graph with vertex set $G$, edge set
$\{(g,gh);g\in G, h\in S_+\}$ with the edge directed from $g$ to $gh$ with color assigned by $h$
(when $h=h^{-1}$ we conventionally draw an undirected edge).  \bigskip

Notice that a Cayley graph in addition to being a \emph{regular} graph, it is also
\emph{vertex-transitive}---\ie all sites are equivalent, in the sense that the graph automorphism
group acts transitively upon its vertices. The Cayley graph is also called \emph{arc-transitive}
when its group of automorphisms acts transitively not only on its vertices but also on its directed
edges.

\subsubsection{The locality principle}
Locality corresponds to require that the evolution is completely determined by a rule involving a
finite number of systems. This means having each system interacting with a finite number of
systems (\ie $|N|<\infty$ in H\ref{i:N}), and having the set of loops generating $F$ as finite and containing only finite
loops. This corresponds to the fact that the group $G$ is {\em finitely presented}, namely both
$S_+$ and $W$ are finite in $G=\<S_+|W\>$.

The quantum walk then corresponds to a unitary operator over the Hilbert space $\sH=\ell^2(G)\otimes
\Cmplx^s$ of the form
\begin{equation}\label{eq:noniso}
  A=\sum_{h\in S} T_h\otimes A_h,
\end{equation}
where $T$ is the {\em right-regular} representation of $G$ on $\ell^2(G)$,
$T_g|g'\>=|g'g^{-1}\>$.

\subsubsection{The isotropy principle}
The requirement of \emph{isotropy} corresponds to the statement that all directions on
$\Gamma(G,S_+)$ are equivalent. Technically the principle affirms that there exists a choice of
$S_+$, a group $L$ of graph automorphisms on $\Gamma(G,S_+)$ that is transitive over $S_+$ and with
faithful unitary (generally projective) representation $U$ over $\Cmplx^s$, such that the
following covariance condition holds
\begin{equation}\label{eq:covW}
A=\sum_{h\in S} T_h\otimes A_h=\sum_{h\in S} T_{l(h)}\otimes U_lA_hU^\dag_l,\quad \forall l\in L.
\end{equation}
As a consequence of the linear independence of the generators $T_h$ of the right regular
representation of $G$ one has that the above condition (\ref{eq:covW}) implies
\begin{align}\label{e:iso2}
A_{l(h^{\pm1})}=U_l A_{h^{\pm1}} U_l^\dag.
\end{align}
Eq. (\ref{e:iso2}) implies that the principle of isotropy requires the Cayley graph $\Gamma(G,S_+)$
to be arc-transitive (see Subsect. \ref{s:homo}).

We remind that the split $S=S_+\cup S_-$ is non unique (and in addition one may add to $S$ the
identity element $e$ corresponding to zero-length loops on each element corresponding to
self-interactions).  Therefore, generally the quantum walk on the Cayley graph $\Gamma(G,S_+)$
satisfies isotropy only for some choices of the set $S_+$. It happens that for the known cases
satisfying all principles along with the restriction to quasi isometric embeddability of $G$ in
Euclidean space (see Subsect. \ref{s:spacetime}) such choice is unique.

\subsubsection{The unitarity principle}
The requirement that the evolution be unitary translates into the following set of equations
bilinear in the transition matrices as unknown 
\begin{align}
\sum_{h\in S}A^\dag_h A_h =\sum_{h\in S}A_h A^\dag_h=I_s,\qquad
\sum_{\shortstack{$\scriptstyle h,h'\in S$\\ $\scriptstyle
      h^{-1}h'=h''$}} A^\dag_h A_{h'}=\sum_{\shortstack{$\scriptstyle
      h,h'\in S$\\ $\scriptstyle h'h^{-1}=h''$}} A_{h'} A^\dag_{h}=0.
      \label{eq:unitarity}
\end{align}
Notice that the structure of equations already satisfy the homogeneity and locality principles. 
The solution of the systems of equations (\ref{eq:unitarity}) is generally a difficult problem.

\subsection{Restriction to Euclidean emergent space}\label{s:spacetime}

How a discrete quantum algorithm on a graph can give rise to a continuum quantum field theory on
space-time? We remind that the flow of the quantum state occurs on a Cayley graph and the evolution
occurs in discrete steps. Therefore the Cayley graph must play the role of a discretized space, whereas the
steps play the role of a discretized time, namely the quantum automaton/walk has an inherent
Cartesian-product structure of space-time, corresponding to a particular chosen observer. We will
then need a procedure for recovering the emergent space-time and a re-interpretation of the notion
of inertial frame and of boost in the discrete, in order to recover Poincar\'e covariance and the
Minkowski structure. The route for such procedure is opened by {\em geometric group theory}, 
a field in pure mathematics initiated by Mikhail Gromov at the beginning of the
nineteen.\footnote{The absence of the appropriate mathematics was the reason of the lack of
  consideration of a discrete structure of space-time in earlier times. Einstein himself was
  considering this possibility and lamented such lack of mathematics. Here a passage reported by
  John Stachel \cite{colodny1986quarks}
\begin{displayquote}{\em
But you have correctly grasped the drawback that the continuum brings. If the molecular view of
matter is the correct (appropriate) one, \ie, if a part of the universe is to be represented by a
finite number of moving points, then the continuum of the present theory contains too great a
manifold of possibilities. I also believe that this too great is responsible for the fact that our
present means of description miscarry with the quantum theory. The problem seems to me how one can
formulate statements about a discontinuum without calling upon a continuum (space-time) as an aid;
the latter should be banned from the theory as a supplementary construction not justified by the
essence of the problem, which corresponds to nothing ``real''. But we still lack the mathematical
structure unfortunately. How much have I already plagued myself in this way!} 
\end{displayquote}} The founding idea is the notion of \emph{quasi-isometric embedding}, which
allows us to compare spaces with very different metrics, as for the cases of continuum and discrete.
Clearly an isometric embedding of a space with a discrete metric (as for the word metric of the Cayley
graph) within a space with a continuum metric (as for a Riemaniann manifold) is not possible. However,
what Gromov realized to be geometrically relevant is the feature that the discrepancy between the
two different metrics is uniformly bounded over the spaces. More precisely, one introduces the
following notion of {\em quasi-isometry}.

\paragraph{Quasi-isometry.} Given two metric spaces $(M_1,d_1)$ and $(M_2,d_2)$, with metric $d_1$ and $d_2$, respectively, a
map $f:(M_1,d_1)\rightarrow (M_2,d_2)$ is a quasi-isometry if there exist constants $A\geq 1$,
$B,C\geq 0$, such that $\forall g_1,g_2\in M_1$ one has
\begin{equation}
\frac{1}{A}d_1(g_1,g_2)-B\leq d_2(f(g_1),f(g_2))\leq A d_1(g_1,g_2)+B,
\end{equation}
and $\forall m\in M_2$ there exists $g\in M_1$ such that
\begin{align}\label{e:quasionto}
d_2(f(g),m)\leq C.
\end{align}
The condition in Eq. (\ref{e:quasionto}) is also called {\em quasi-onto}. 

\medskip
It is easy to see that quasi-isometry is an equivalence relation. It can also be proved that the quasi-isometric class is
an invariant of the group, \ie it does not depend on the presentation, \ie  on the Cayley graph.
Moreover, it is particularly interesting for us that for finitely generated groups, the
quasi-isometry class always contains a smooth Riemaniann manifold \cite{harpe}. Therefore, for a
given Cayley graph there always exists a Riemaniann manifold in which it can be quasi-isometrically
embedded, which is unique modulo quasi-isometries, and which depends only on the group $G$ of the
Cayley graph.  Two examples are graphically represented in Fig. \ref{fig:cayley}.  

\subsubsection{Geometric group theory}
With the idea of quasi-isometric embedding, geometric group theory connects geometric properties of the embedding Riemaniann
spaces with algebraic properties of the groups, opening the route to a {\em geometrization of group
  theory}, including the generally hard problem of establishing properties of a group that is given by
presentation only.\footnote{One should consider that the Dehn's problem of establishing if two
  words of generators correspond to the same group element is generally undecidable. The same is
  true for the problem of establishing if the presentation corresponds to the trivial group!}
\begin{figure}[h!]
  \begin{center}
    \includegraphics[width=\textwidth]{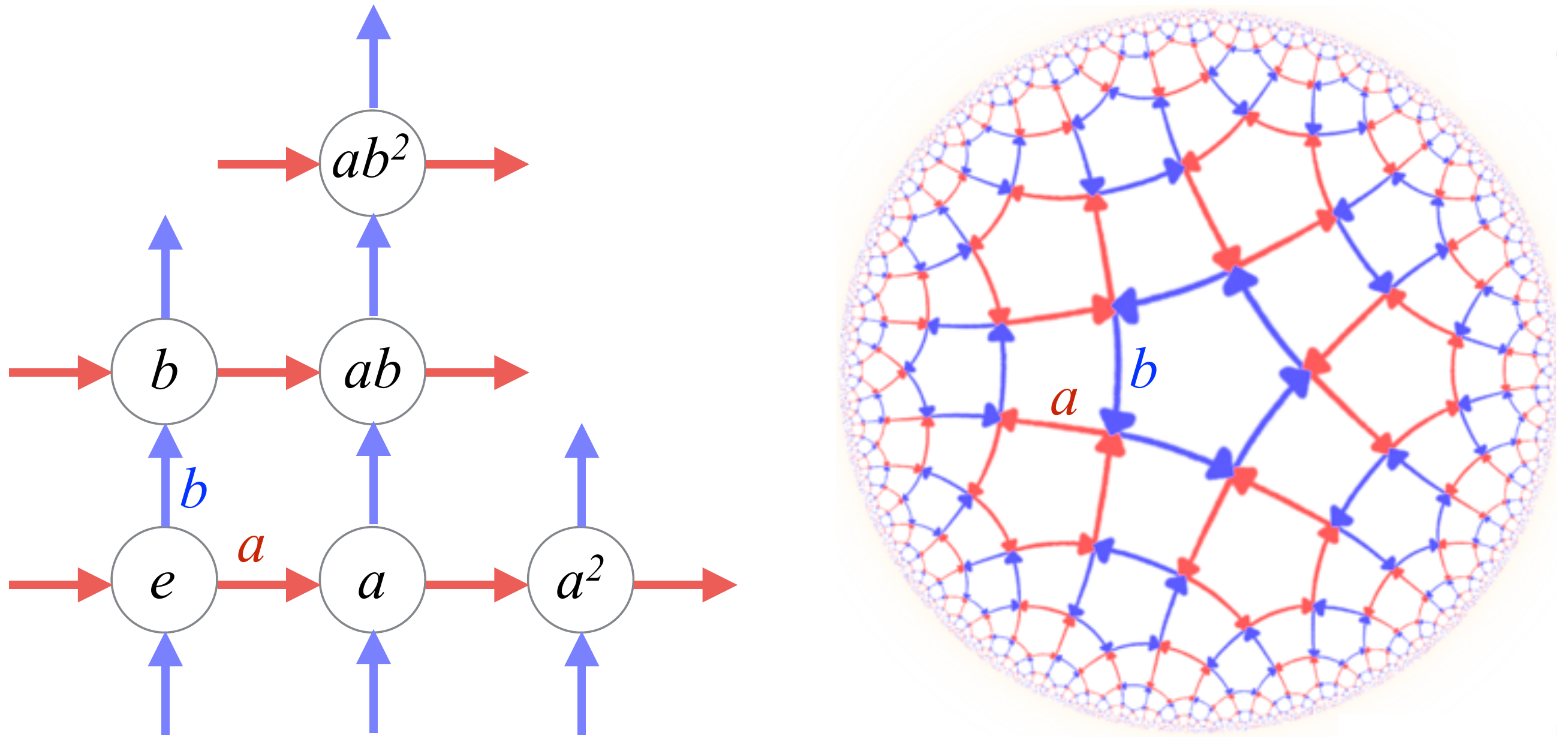}
    \caption{From  Ref. \cite{FOPDP16}.  (Colors online). Given a group $G$ and a set $S_+$ of generators, the Cayley graph
      $\Gamma(G,S_+)$ is defined as the colored directed graph having set of nodes $G$, set of edges
      $\{(g,gh); g \in G, h \in S_+\}$, and a color assigned to each generator $h \in S_+$. Left figure: 
      the Cayley graph of the Abelian group $\Z^2$ with presentation
      $\Z^2=\langle a,b|aba^{-1}b^{-1}\rangle$, where a and b are two commuting generators.
      Right figure: the Cayley graph of the non-Abelian group $G=\langle a,b|a^5,b^5,(ab)^2\rangle$.
      The Abelian-group graph is embedded into the Euclidean space $\Reals^2$, the non-Abelian
      $G$ into the Hyperbolic space $\mathbb{H}_2$ with negative curvature. \label{fig:cayley}}
  \end{center}
\end{figure}

The possible groups $G$ that are selected from our principles are infinitely many, and we need to
restrict this set to start the search for solutions of the unitarity conditions (\ref{s:spacetime})
under the isotropy constraint. Since we are interested in a theory involving infinitely many systems
(we take the world as infinite!), we will consider {\em infinite} groups only. This means that when
we consider an Abelian group, we always take it as {\em free}, namely its only relators are those
establishing the Abelianity of the group. This is the case of  $G=\Z^d$, with $d\geq 1$.

A paradigmatic result \cite{harpe} of geometric group theory is that an infinite group
$G$ is quasi-isometric to an Euclidean space $\Reals^d$ if and only if $G$ is \emph{virtually-Abelian}, 
namely it has an Abelian subgroup $G'\subset G$ isomorphic to $\Z^d$ of
finite index (namely with a finite number of cosets). Another result is that a group has
polinomial growth iff it is \emph{virtually-nihilpotent}, and if it has exponential growth then it not 
virtually-nihilpotent, and in particular non Abelian, and is quasi-isometrically embeddable in a manifold
with negative curvature.

In the following we will restrict to groups that are quasi-isometrically embeddable in Euclidean
spaces. As we will see soon, such restriction will indeed lead us to free quantum field theory in
Euclidean space. It would be very interesting to address also the case of curved spaces, to get
hints about quantum field theory in curved space. Unfortunately, the case of negative curvature
corresponds to groups, as the Fuchsian group in Fig.  \ref{fig:cayley}, whose unitary
representations (that we need here) are still unknown \cite{Farb,Drutu,Tessera}. The
virtually-nihilpotent case also would be interesting, since it corresponds to a Riemaniann manifold
with variable curvature \cite{Tessera}, however, a Cayley graph that can satisfy the isotropy
constraint could not be found yet \cite{unpubDEP}.

\bigskip I close this section with some comments about the remarkable closeness in spirit between
the present program and the geometric group theory program. The main general goal of geometric group theory is the {\em geometrization of group theory}, which is achieved studying finitely-generated groups $G$ as symmetry groups of
metric spaces $X$, with the aim of establish connections between the algebraic structure of $G$ with
the geometric properties of $X$ \cite{Kapovich-notes}. In a specular way the present program is an
{\em algorithmization of theoretical physics}, with the general goal of deriving QFT (and ultimately
the whole physics) from quantum algorithms with finite complexity, upon connecting the algebraic
properties of the algorithm with the dynamical features of the physical theory. This will allow a
coherent unified axiomatization of physics without physical primitives, preparing a logically
coherent framework for a theory of quantum gravity.

\section{Quantum Walks on Abelian groups and free QFT as their relativistic regime}\label{s:Abelian}
As seen in Subsect. \ref{s:spacetime}, from the huge and yet mathematically unexplored set of
possibilities for the group $G$ of the quantum walk, we restrict to the case of $G$
virtually-Abelian, which corresponds to $G$ quasi-isometrically embeddable in an Euclidean space.
As we will see in the present section, the free QFT that will be derived from such choice exactly corresponds to the known
QFT in Euclidean space.

Since we are interested in the physics occurring in $\Reals^3$, we need to classify all possible
Cayley graphs of $G$ having $\Z^3$ as subgroup with finite index, and then select all graphs that
allow the quantum walk to satisfy the conditions of isotropy and unitarity. We can proceed by
considering increasingly large dimension $s>0$ (defined in H\ref{i:s}), which ultimately 
corresponds to the dimension of the field--e~.g. a scalar field for $s=1$, a spinor field for $s=2$,
etc.

\subsection{Induced representation, and reduction from virtually Abelian to Abelian quantum walks}
An easy way to classify all quantum walks on Cayley graphs with virtually Abelian groups is provided
by a theorem in Ref.  \cite{d2015virtually}, which establishes the following

\medskip{\em A quantum walk on the Cayley graph of a virtually Abelian group $G$ with Abelian
  subgroup $H\subset G$ of finite index $i_H$ and dimension $s$ is also a quantum walk on the Cayley
  graph of $H$ with dimension $s'=si_H$.  }\medskip

\begin{figure}
\begin{center}
\includegraphics[width=.8\textwidth]{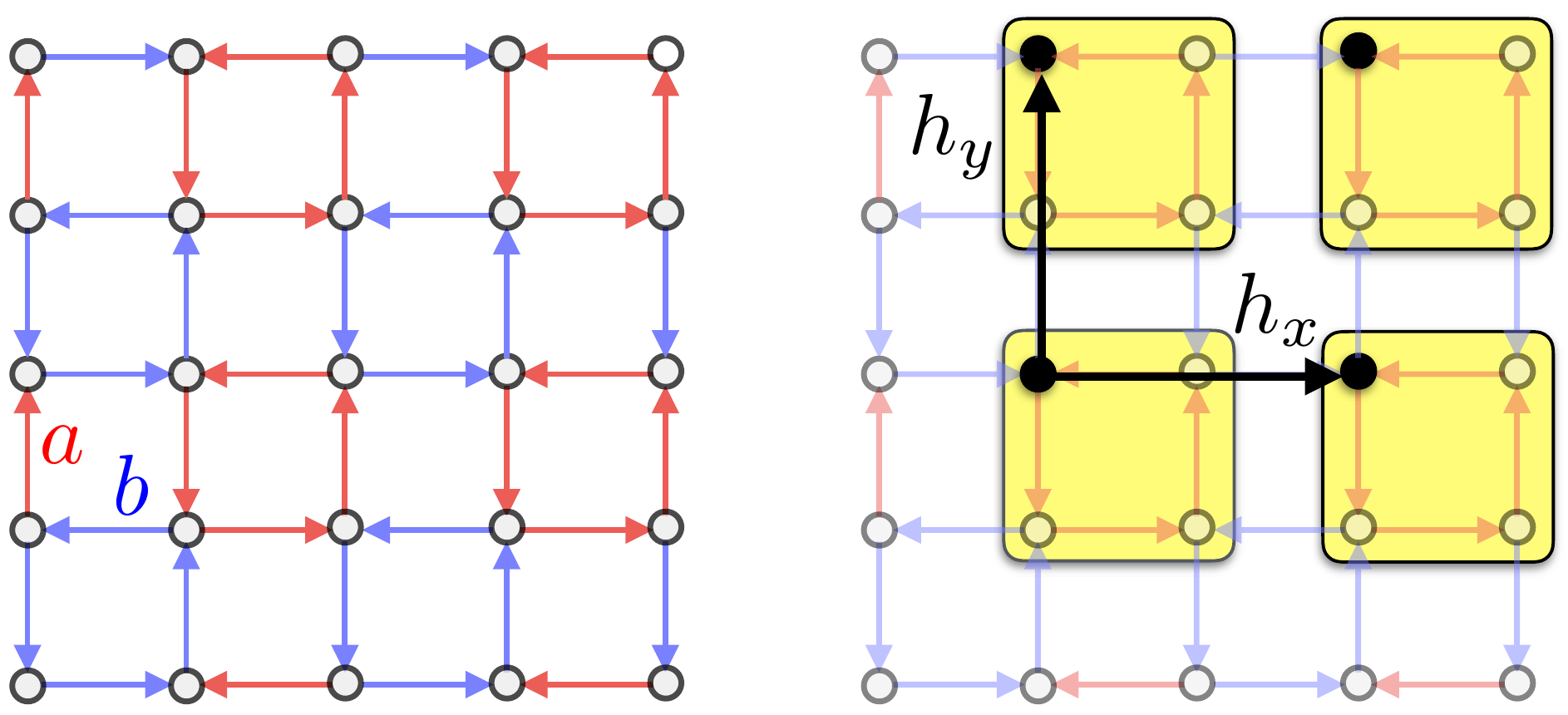}
\includegraphics[width=.8\textwidth]{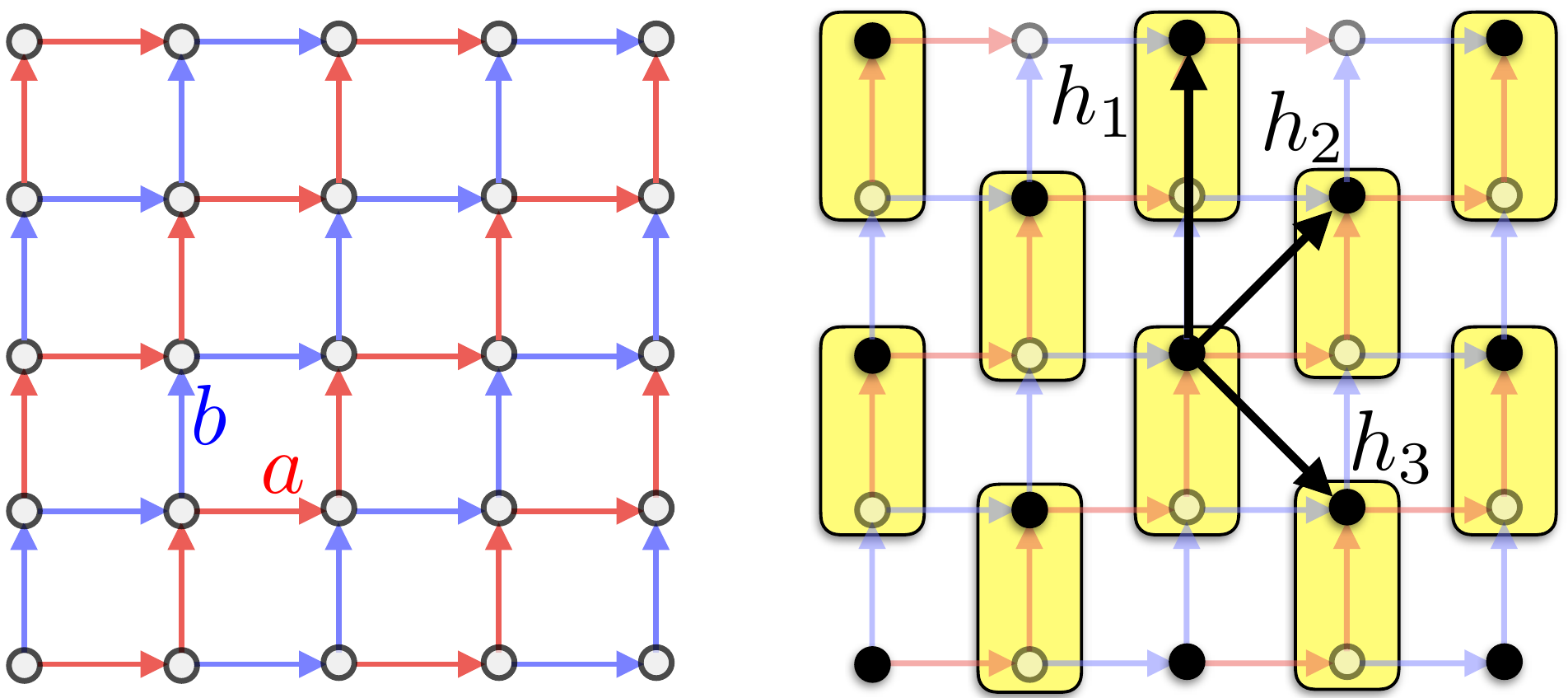}
\caption{From Ref. \cite{d2015virtually}. (Colors online). Two examples of reduction of a quantum walk
  on the Cayley graph of a virtually Abelian group $G$ to that of a quantum walk on the Cayley graph
  of an Abelian subgroup $H\subset G$ with finite index $i_H$. The graphs on the left of the figures
  are the Cayley graph of $G$ (it is easy to see that both groups are non Abelian).  The graphs on
  the right represents a choice of the Cayley graph of the subgroup $H=\Z^2$, with the tiling
  corresponding to the induced representation (the elements of $H$ are the black bullets). Top figures:
  $G=\langle a,b\ |\ a^4,b^4,{\left(ab\right)}^{2}\rangle $. The index is $i_H=4$. The subgroup
  generators are $h_x=a^{-1}b$ and $h_y=ba^{-1}$. The tiling is defined by the coset representatives
  $e,a,a^2,a^3$.  Bottom figures: $G= \langle a,b\ |\ a^2b^{-2}\rangle $. The index is $i_H=2$. The subgroup
  generators are $h_1=ba$ and $h_2=a^2$ (or $h_1=ba$ and $h_3=ab^{-1}$), with the tiling the cosets
  representatives $e,a$.\label{fig:example}}
\end{center}
\end{figure}

This is just the {\em induced-representation} theorem
\cite{mackey1951induced,mackey1952induced,mackey1953induced} in group theory, here applied to quantum
walks.  The multiple dimension $s'=si_h$ corresponds to tiling the Cayley graph of $G$ with a tile
made with a particular choice of the cosets of $H$.  The new set of transition matrices of the new
walk for $H$ can be straightforwardly evaluated in terms of those for $G$ (generally self
interactions within the same tile can occur, corresponding to zero-length loops in the Cayley graph).
In Fig.  \ref{fig:example} two examples of such tiling procedure are given.

The induced-representation method guarantees that scanning all possible virtually Abelian quantum
walks for increasing $s$ is equivalent to scan all possible Abelian quantum walks, since \eg the set
of Abelian walks of dimension $s=nm$ will contain all virtually Abelian walks with $s=n$ and index
$m$, etc. We therefore resort to consider only Abelian groups.

\subsection{Isotropy and orthogonal embedding in $\Reals^3$}\label{s:referback}
We will also assume that the representation of the isotropy group
$L$ in \eqref{eq:covW} induced by the embedding in $\Reals^3$ is orthogonal, which implies that the
graph-neighborhood is embedded in a sphere $S^2\subset \Reals^3$ (we want homogeneity and isotropy
to hold locally also in the embedding space $\Reals^3$). We are then left with the classification of
the Cayley graphs of $\Z^3$ satisfying the isotropic embedding in $\Reals^3$: these are just the
Bravais lattices.

\subsection{Quantum Walks with Abelian $G$}\label{s:WeylQW}
When $G$ is Abelian we can greatly simplify the study of the quantum walk by using the
wave-vector representation, based on the fact that the irreducible representations of $G$ are
one-dimensional.  The interesting case is for $d=3$, but what follows holds for any dimension $d$.
We will label the group elements by vectors $\bg\in\Z^d$, and use the additive notation for
the group composition, whereas the right-regular representation of $\Z^d$ on
$\ell_2(\Z^d)$ will be written as $T_{\bh}|\bvec g\>=|\bg-\bh\>$. This can be diagonalized
by Fourier transform, corresponding to write the operator $A$ in block-form in terms of the
following direct-integral
\begin{align}
  \label{eq:weylautomata} 
  A = \int_B\operatorname d^3 \! \bk  \,  |{\bk}\>\< {\bk}| \otimes
  A_{\bk},\qquad A_\bk:=\sum_{\bh\in S}e^{-i\bk\cdot\bh}A_\bh,\quad |\bk\>:=\frac{1}{\sqrt{|B|}}\sum_{\bg\in G}e^{-i\bk\cdot\bg}|\bg\>.
\end{align}
where $B$ is the {\em Brillouin zone}, and $|\bk\>$ is a {\em plane wave}.\footnote{The Brillouin
  zone is a compact subset of $\Reals^3$ corresponding to the smallest region containing only
  inequivalent wave-vectors $\bk$. (See Ref. \cite{PhysRevA.90.062106} for the analytical expression.)}  
 Notice that the quantum walk is unitary if and only if $A_\bk$ is unitary for every $\bk\in B$.

\subsection{Dispersion relation}
The spectrum $\{e^{-i\omega^{(i)}_\bk}\}$ of the operator $ A_\bk$ is usually given in terms of the
so-called {\em dispersion relations} $\omega^{(i)}_\bk$ versus $\bk$. As in usual
wave-mechanics, the speed of the wave-front of a plane wave is given by the {\em phase-velocity}
$\omega^{(i)}_\bk/|\bk|$, whereas the speed of a narrow-band packet peaked around the value
wave-vector $\bk_0$ is given by the {\em group velocity} $\nabla_\bk\omega^{(i)}_\bk$ evaluated at
$\bk_0$. 

\subsection{The relativistic regime}\label{ss:limit}
As we will see in Sect. \ref{s:phydim} an heuristic argument will lead us to set the scale of
discreteness of the quantum walk (and similarly the quantum cellular automaton for the interacting
theory) at the Planck scale. The domain $|\bk|\ll1$ then corresponds to wave-vectors much smaller
than the Planck vector, which is much higher than any ever observed wave-vector.\footnote{The 
highest momentum observed is that of a ultra-high-energy cosmic ray, which is $k\sim10^{-8}.$} 
Such regime includes that of usual 
particle physics, and is called {\em relativistic regime}. To be precise, the regime is defined by a
set of wavepackets that are peacked around $\bk=0$ with r.m.s.  value much smaller than the Planck
wave-vector, which we will refer shortly to as {\em narrow-band wave-packets}.

I want to emphasize here that we have never used any mechanical concept in our derivation of the
quantum walk, including the notion of Hamiltonian: the dynamics is given in term of a
single unitary operator $A$. A notion of effective Hamiltonian could be considered as the logarithm
of $A$, which would correspond to an Hamiltonian providing the same unitary evolution, and which
would even interpolate it between contiguous steps. For this reason we will call such an operator
{\em interpolating Hamiltonian}. In the Fourier direct-integral representation of the operator, the
interpolating Hamiltonian will be given by the identity $e^{-i H(\bk)}:=A_\bk$. It is easy to see
that the relativistic limit $H_0(\bk)$ of $H(\bk)$, corresponding to consider narrow-band
wave-packets centered at $\bk=0$, is achieved by expanding it at the first order in $|\bk|$, \ie 
$H(\bk)=H_0(\bk)+\mathcal O(|\bk|^2)$. The interpolated continuum-time evolution in 
the relativistic regime will be then given by the first-order differential equation in the 
Schr\"{o}dinger form
\begin{equation}\label{eq:diff}
  i\partial_t\psi(\bk,t)=H_0(\bk)\psi(\bk,t).
\end{equation}
Rigorous quantitative approaches to judge the closeness between free QFT and the relativistic
regime of the quantum walk have been provided in Ref.~\cite{Bisio2015244} in terms of channel
discrimination probability, and in Ref. ~\cite{PhysRevA.90.062106} in terms of fidelity between the
two evolutions. Numerical values will be provided at the end of Subsect. \ref{s:dirac}.

\subsection{Schr\"{o}dinger equation for the ultra-relativistic regime}
\begin{figure*}
  \centering\def\sizeHerm{0.44}
\includegraphics[width=\sizeHerm\textwidth]{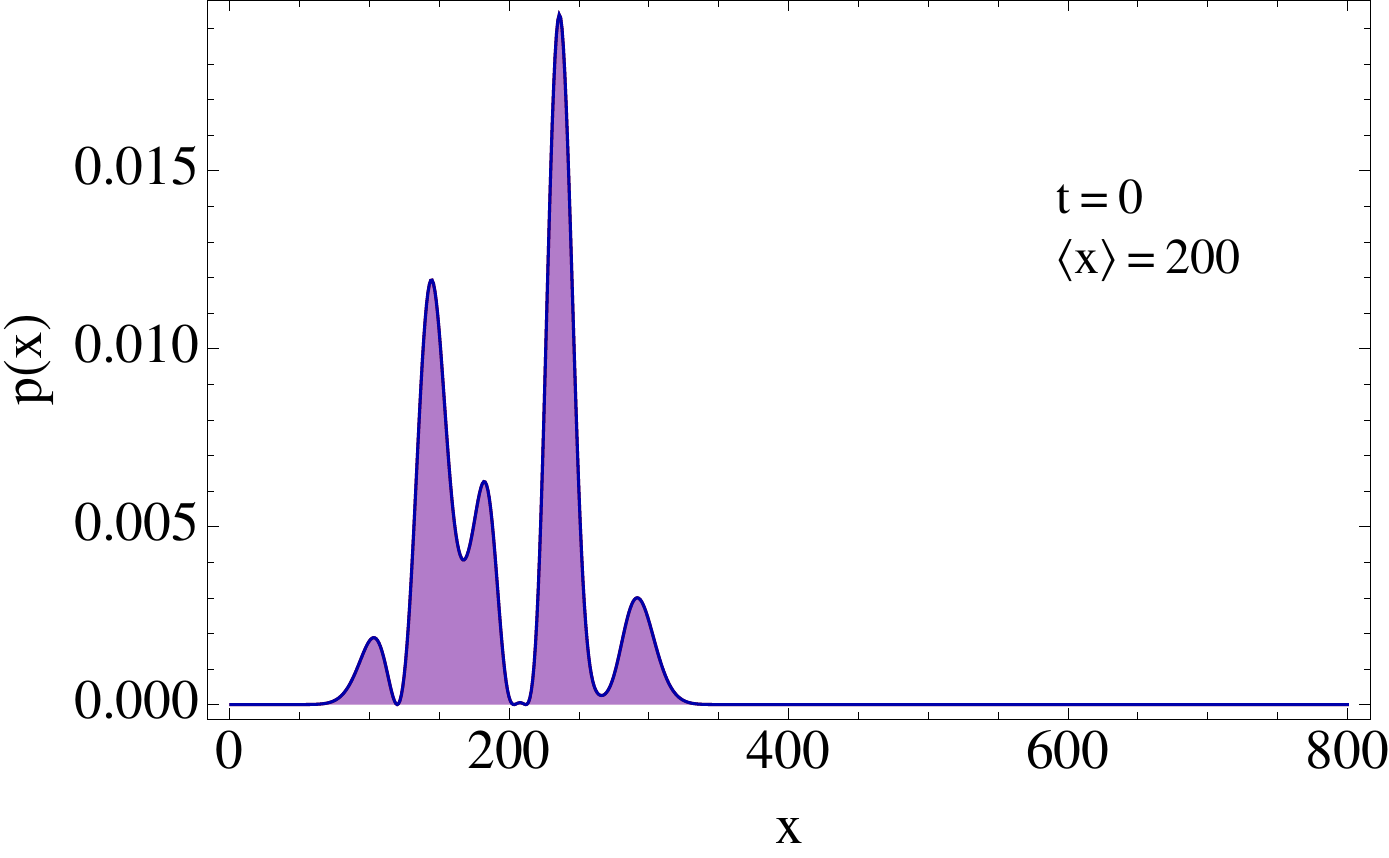}
\qquad \qquad 
\includegraphics[width=\sizeHerm\textwidth]{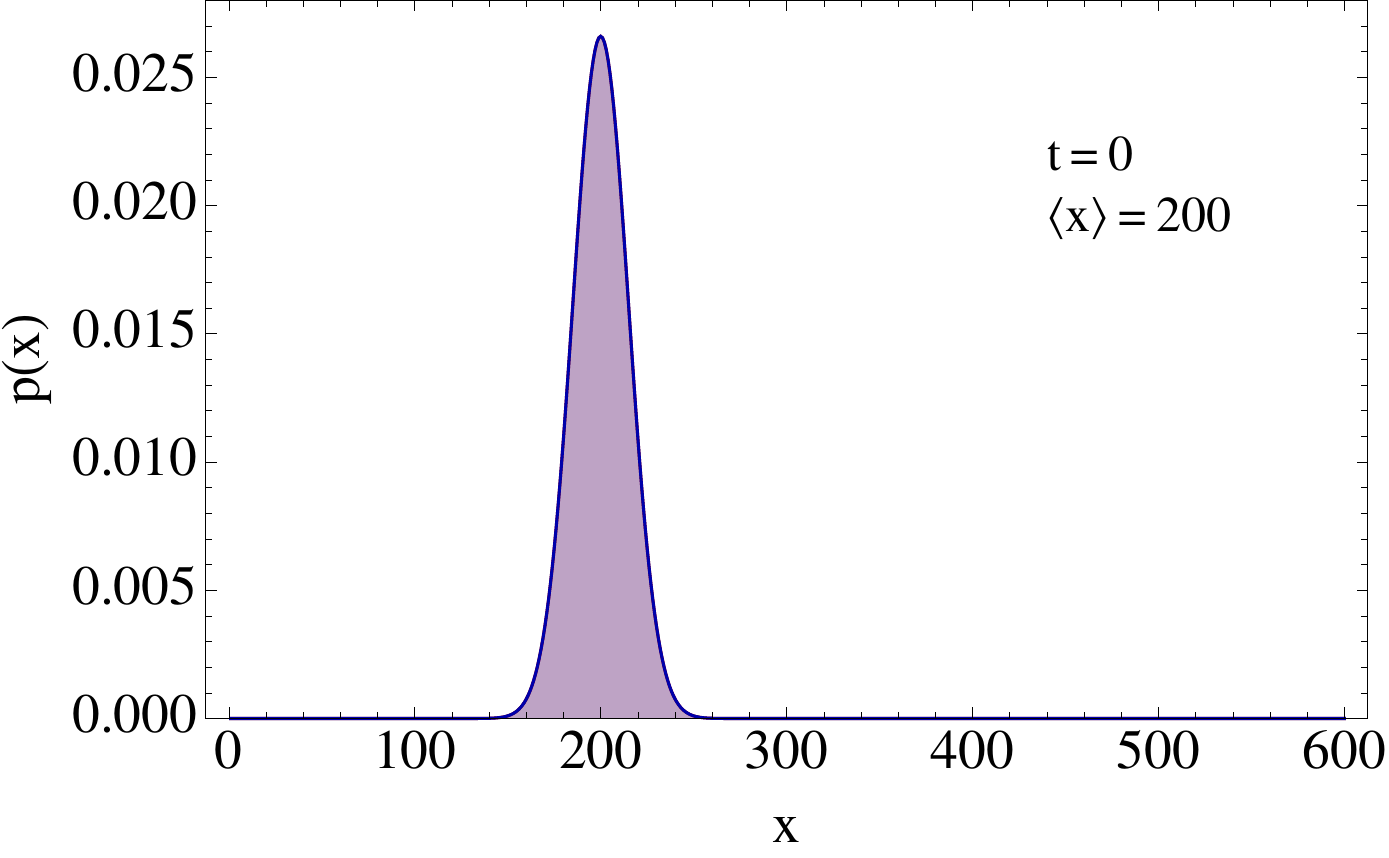}\\
\includegraphics[width=\sizeHerm\textwidth]{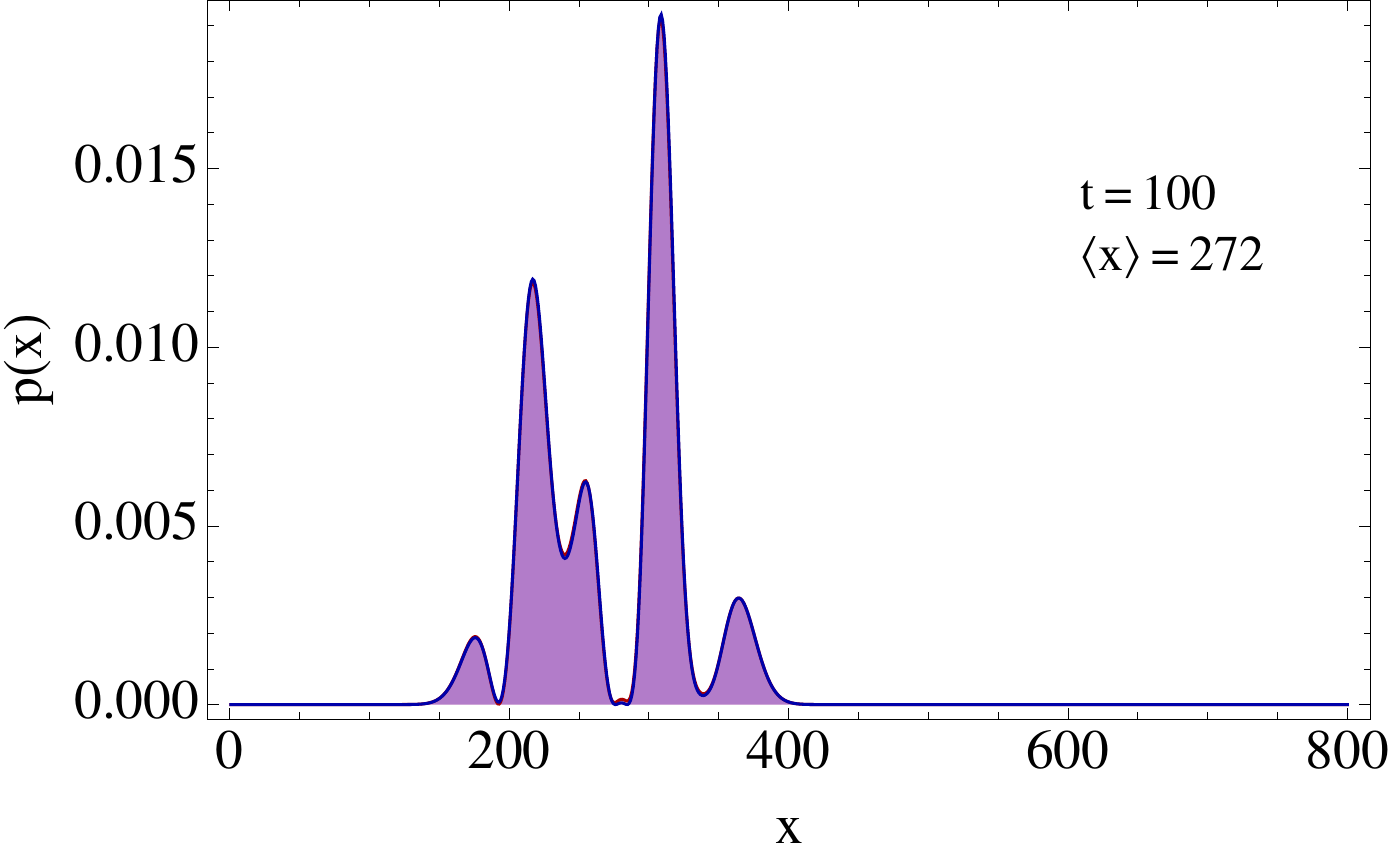}
\qquad \qquad 
\includegraphics[width=\sizeHerm\textwidth]{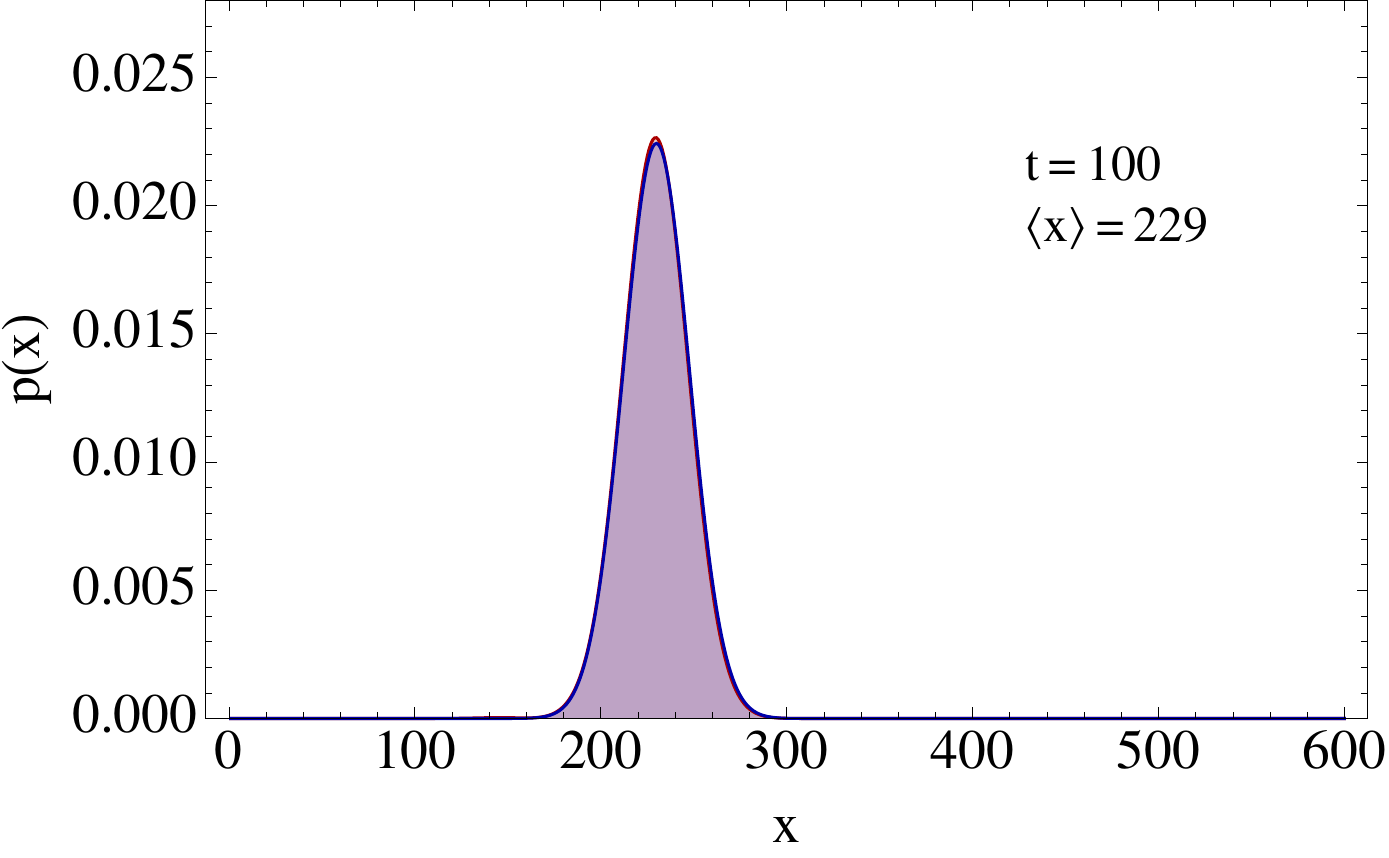}\\
\includegraphics[width=\sizeHerm\textwidth]{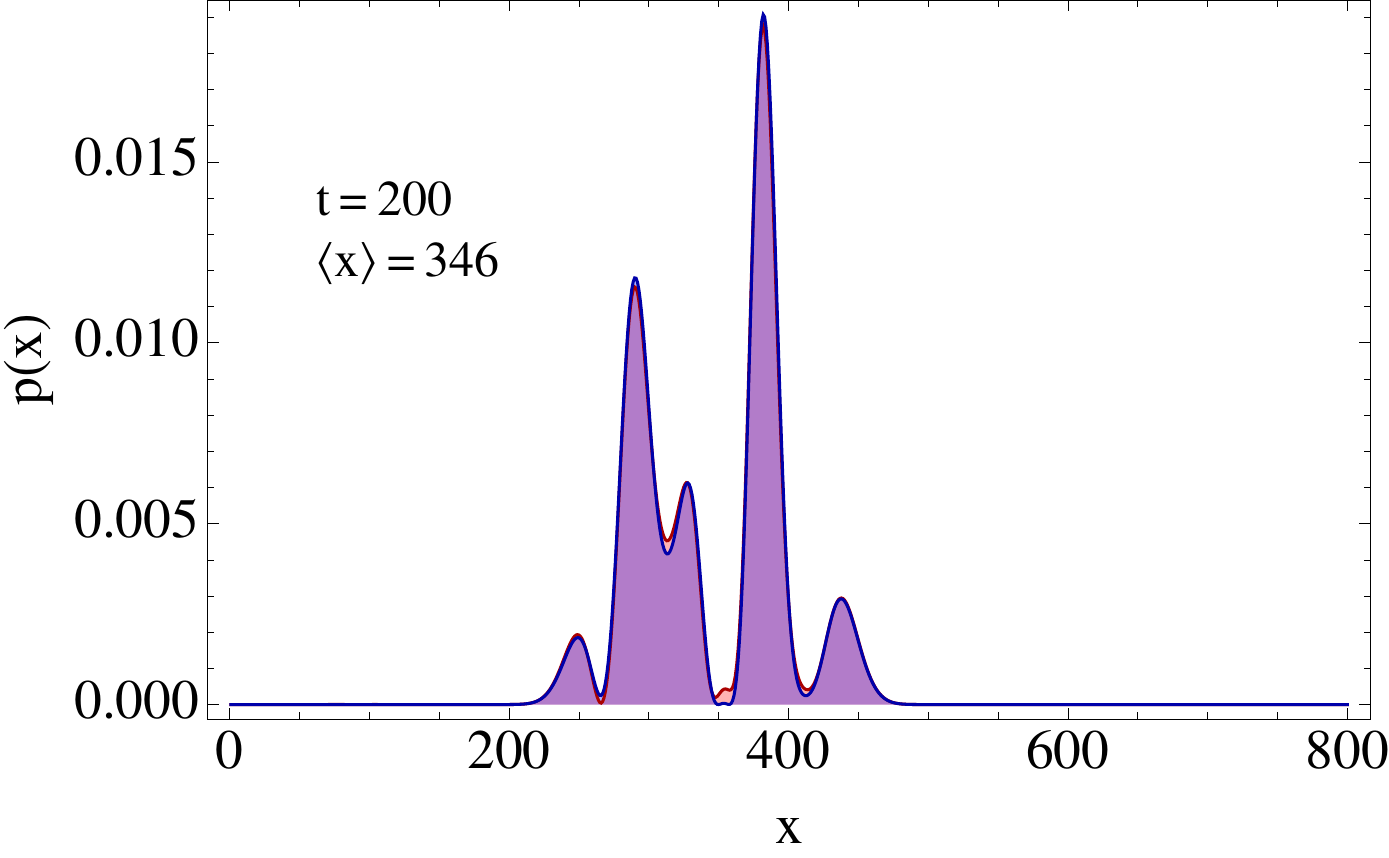}
\qquad \qquad 
\includegraphics[width=\sizeHerm\textwidth]{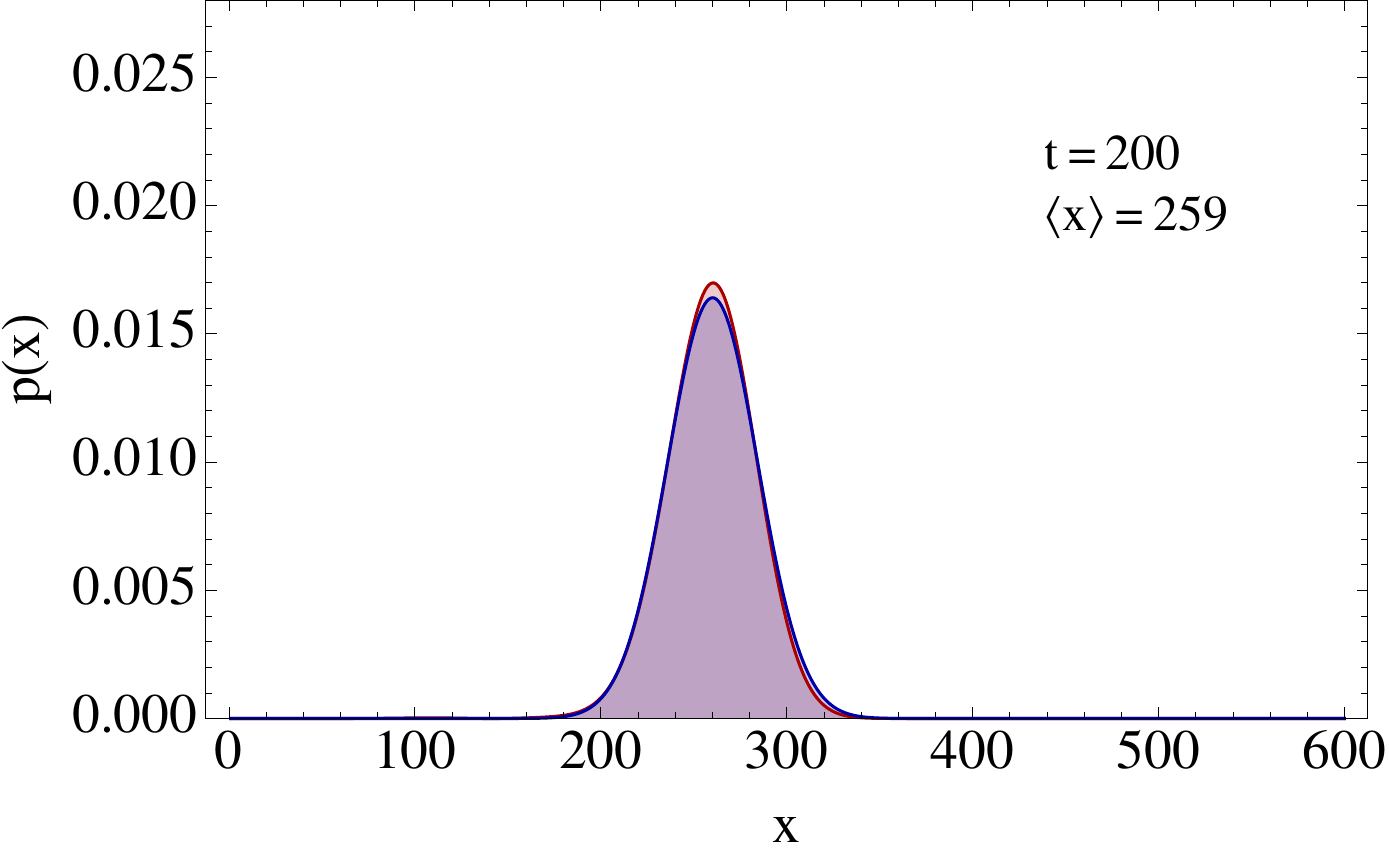}\\
\includegraphics[width=\sizeHerm\textwidth]{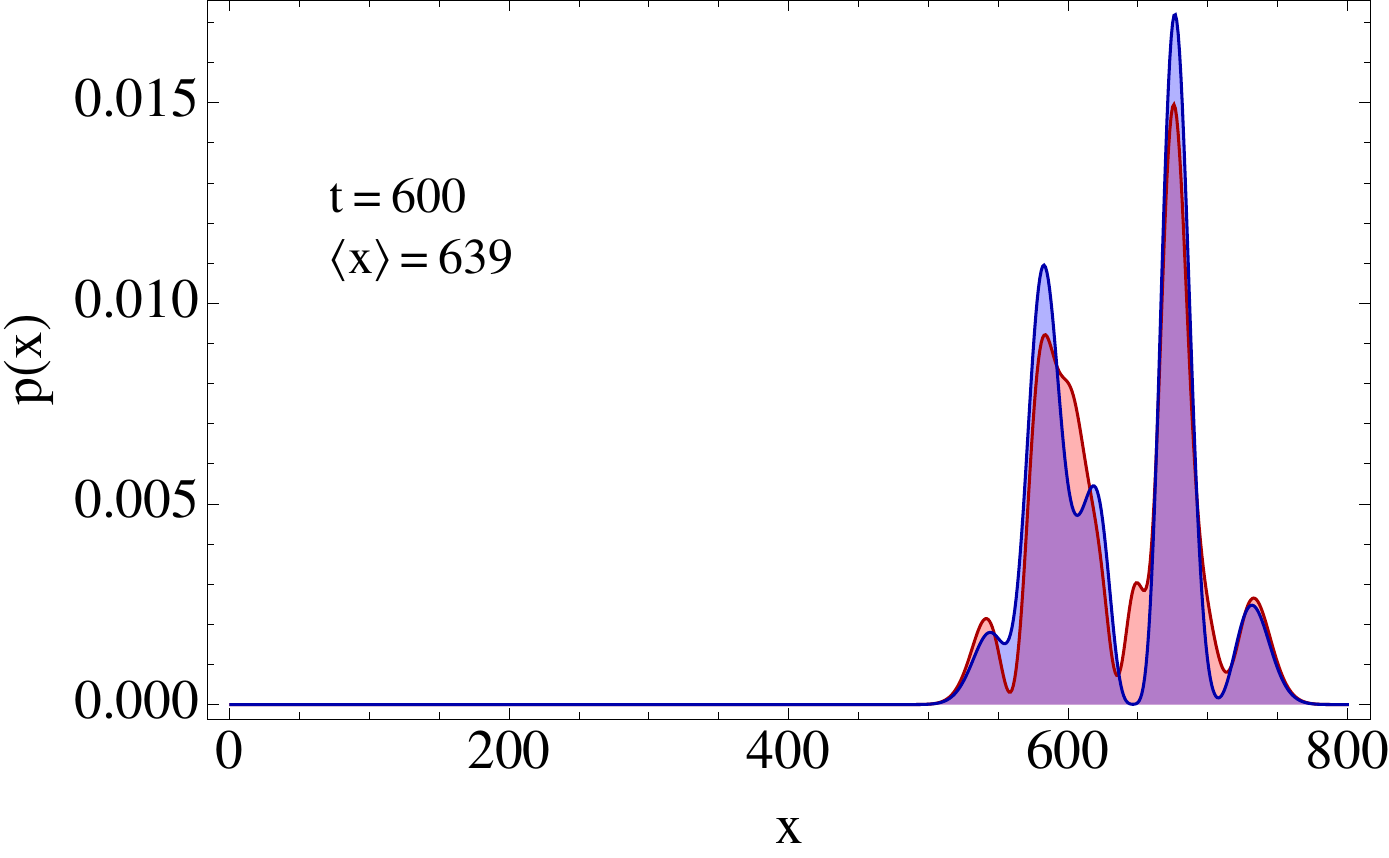}
\qquad \qquad 
\includegraphics[width=\sizeHerm\textwidth]{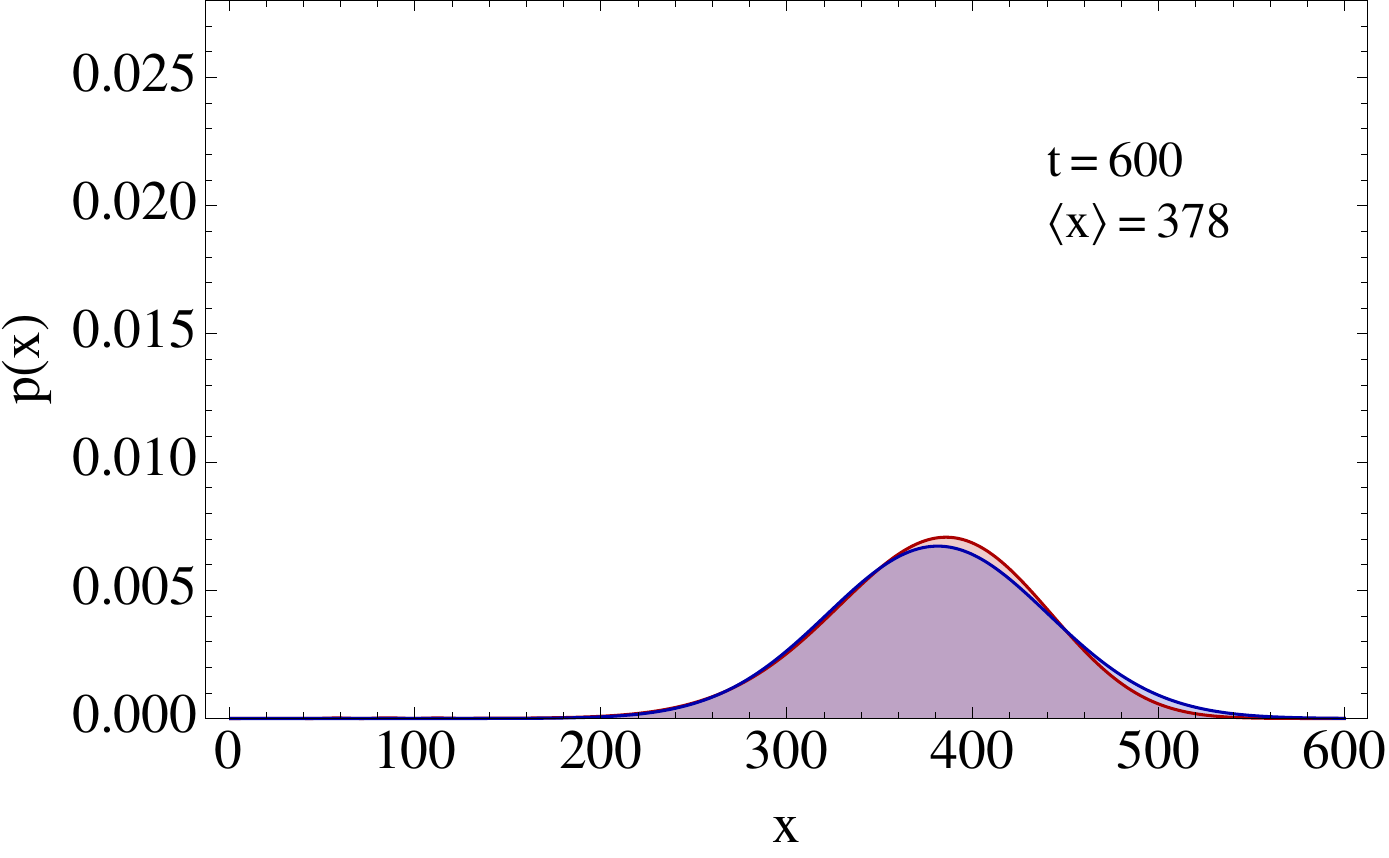}
\caption{From Ref. \cite{Bisio2015244}. (Colors online). Test of the quality of the approximation of
  the Schr\"odinger equation (\ref{eq:approxstate}) at for different time $t$ of the Dirac quantum
  walk with mass $m=0.6$ in one space dimension of Ref.\cite{Bisio2015244}. Comparison of the
  probability distribution (in red) and the solution of the Schr\"odinger equation (in blue).  Right figures: 
  the state is a superposition of Hermite functions multiplied by the Gaussian
  peaked around momentum $k_0=3\pi/10$, for drift and diffusion coefficients $v=0.73$ and $D=0.31$,
  respectively.  The mean value moves at the group velocity given by the drift coefficient $v$. The
  approximation remains accurate even for position spread $\hat\sigma=20$ Planck lengths. Left
    figures: The same four times comparison for the quantum walk with $m=0.4$, and an initial
  Gaussian state peaked around the momentum $k_0=0.1$.  In this case the drift velocity and the
  diffusion coefficient are respectively $v=0.22$ and $D=2.30$.\label{fig:Hermite}}
\end{figure*}

In the ultra-relativistic regime of wave-vectors comparable to the Planck vector, an obvious
option is that of evaluating the evolution by a
numerical evaluation of the exact quantum walk.\footnote{A fast numerical technique to evaluate the
  quantum walk evolution numerically exploits the Fourier transform. For an application to the Dirac
  quantum walk see Ref. \cite{d2016discrete}.} However, even in such regime we still have an
analytical method available for evaluating the evolution of some common physical states. Indeed, 
for narrow-band wave packets centered around any value $\bk_0$ one can write a dispersive 
Schr\"odinger equation by expanding the interpolating Hamiltonian $H(\bk)$
around $\bk_0$ at the second order, thus obtaining 
\begin{equation}\label{eq:approxstate}
i\partial_t \tilde\psi(\bvec x,t)=\pm[\bvec v\cdot\bvec \nabla+\tfrac{1}{2}\bvec D\cdot\bvec
\nabla\bvec \nabla]\tilde\psi(\bvec x,t),
\end{equation}
where $\tilde\psi(\bvec x,t)$ is the Fourier transform of $\tilde\psi(\bk,t):=e^{-i\bk_0\cdot\bvec
  x+i\omega_0 t}\psi(\bk,t)$, $\bvec v=\left(\bvec\nabla_{\bk}\omega\right)(\bk_0)$ is the drift
vector, and $\bvec D=\left(\bvec\nabla_{\bk}\bvec\nabla_{\bk}\omega\right)(\bk_0)$ is the diffusion
tensor. This equation approximates very well the evolution, even in the Planck regime and for large
numbers of steps, depending on the bandwith (see an example in Fig. \ref{fig:Hermite} from Ref.
\cite{Bisio2015244}).

\subsection{Recovering the Weyl equation\protect\footnote{This section is a synthesis of the results of Ref.
    \cite{PhysRevA.90.062106}. It should be noticed that there isotropy is not even assumed in
    solving Eqs.  (\ref{eq:unitarity}). A simplified derivation making use of isotropy and full
    detailed analysis of all possible Cayley graphs will be available soon \cite{unpubDEP}.}}\label{s:weyl}

In Subsect. \ref{s:referback} we were left with the classification of the Cayley graphs of $\Z^3$
satisfying the isotropic embedding in $\Reals^3$, which are just the Bravais lattices. For dimension
$s=1$ it is easy to show that the only solution of the unitarity constraints gives the trivial
quantum walk $A=I$.\footnote{Also more generally one has $A=T_h$.} 
We then consider $s=2$. Now, the only inequivalent isotropic Cayley graphs are
the primitive cubic (PC) lattice, the body centered cubic (BCC), and the rhombohedral.  However only
in the BCC case, whose presentation of $\Z^3$ involves four vectors
$S_+=\{\bh_1,\bh_2,\bh_3,\bh_4\}$ with relator $\bh_1+\bh_2+\bh_3+\bh_4=0$, one finds solutions
satisfying all the assumptions of Section \ref{s:assumptions}. The isotropy group is given by the
group $L$ of binary rotations around the coordinate axes, with the unitary projective representation
on $\Cmplx^2$ given by $\{I,i\sigma_x,i\sigma_y,i\sigma_z\}$. The group $L$ is transitive on the
four BCC generators of $S_+$. There are only four solutions (modulo unitary conjugation) that can be
divided in two pairs $A^\pm$ and $B^\pm$. The two pairs of solutions are connected by transposition
in the canonical basis, \ie $A_\bk^\pm=(B_\bk^\pm)^T$. The solutions $B_\bk^\pm$ can be also
obtained from the solution $A_\bk^\pm$ by shifting the wave-vector $\bk$ inside the Brillouin
zone\footnote{The first Brillouin zone $B$ for the BCC lattice is defined in Cartesian coordinates
  as $-{\sqrt3}\pi\leq k_i\pm k_j\leq{\sqrt3}\pi, \, i\neq j\in\{x,y,z\}$.} to the vectors \cite{PhysRevA.90.062106}
\begin{equation}\label{e:kkk}
\v{k_1}=\frac{\pi}{2}(1,1,1),\quad \v{k_2}=-\frac{\pi}{2}(1,1,1),\quad \v{k_3}=-\frac{\pi}{2}(1,0,0).
\end{equation}
The $A_\bk^\pm$ solutions in the wave-vector representation are
\begin{equation}\label{eq:weyl3d}
A^\pm_{\bk}=I u^{\pm}_\bk-i\boldsymbol{\sigma}^\pm\cdot
  \tilde\bn^{\pm}_\bk
\end{equation}
with 
\begin{equation}\label{eq:weyl3dxyz}
{\tilde\bn}^{\pm}_{\bk} :=
\begin{pmatrix}
s_x c_y c_z \mp c_x s_y s_z\\
c_x s_y c_z \pm s_x c_y s_z\\
c_x c_y s_z \mp s_x s_y c_z
\end{pmatrix},\quad
u^{\pm}_{\bk} :=  c_x c_y c_z \pm s_x s_y s_z ,\\
\end{equation}
where $c_i:=\cos(k_i/\sqrt3)$, $s_i:=\sin(k_i/\sqrt3)$, and $\boldsymbol{\sigma}^+=\boldsymbol{\sigma}$,
$\boldsymbol{\sigma}^-=\boldsymbol{\sigma}^T$. The spectrum of $A^\pm_\bk$ is
$\{e^{-i\omega^{\pm}}_\bk\}$, with dispersion relation given by
\begin{equation}
\omega^{\pm}_\bk=\arccos(c_xc_yc_z\mp s_xs_ys_z).
\end{equation}
It is easy to get the relativistic limit of the quantum walk using the procedure in Subsect.
\ref{ss:limit}. This simply corresponds to substituting $c_i=1$ and $s_i=k_i/\sqrt{3}$ in Eq.
(\ref{eq:weyl3dxyz}), thus obtaining
\begin{equation}\label{eq:diff}
  i\partial_t\psi(\bk,t)=\frac{1}{\sqrt{3}}\boldsymbol{\sigma}^\pm\cdot\bk\,\psi(\bk,t).
\end{equation}
Eqs. (\ref{eq:diff}) are the two Weyl equations for the left and the right chiralities. For $G=\Z^d$
with $d=1,2$ one obtains the Weyl equations in dimension $d=1,2$, respectively \cite{PhysRevA.90.062106}.
All the three quantum walks have the same form in Eq. (\ref{eq:weyl3d}), namely
\begin{align}\label{Akd}
A_\bk=u_\bk I-i\boldsymbol{\sigma}\cdot\tilde\bn_\bk,
\end{align}
with dispersion relation
\begin{align}\label{e:omega}
\omega_\bk=\arccos{u_\bk},
\end{align}
and with the analytic expression of $u_k$ and $\bn_k$ depending on $d$ and on the chirality (see Ref.
\cite{PhysRevA.90.062106}). Since the quantum walks in Eq. (\ref{eq:weyl3dxyz}) or (\ref{Akd}) have
the Weyl equations as relativistic limit, we will also call them {\em Weyl quantum walks}.

The interpolating Hamiltonian is $H(\bk)=\boldsymbol{\sigma}\cdot\bn_\bk$,
with $\bn_\bk:= (\omega_{\bk}/\sin\omega_{\bk})\tilde{\bn}_{\bk}$ playing the role of an helicity
vector, and with relativistic-limit being given by $H_0(\bk)=\tfrac{1}{\sqrt{d}} \boldsymbol{\sigma}\cdot
\bk$, which coincides with the usual Weyl Hamiltonian in $d$ dimensions upon interpreting the
wave-vector $\bk$ as the particle momentum.

\medskip
We conclude the present subsection by emphasizing that one additional advantages of the discrete framework is that the Feynman path-integral is well defined, and it is also exactly calculated analytically in some cases. Indeed, in Refs. \cite{path1d} and \cite{path2d} the discrete Feynman propagator for the Weyl quantum walk has been analytically evaluated with a closed form for dimensions $d=1$ and $d=2$, and the case of dimension $d=3$ will be published soon \cite{path3d}.

\subsection{Recovering the Dirac equation}\label{s:dirac}
From subsection \ref{s:weyl} we know that all quantum walks derivable from our principles for $s=2$
give the Weyl equation in the relativistic limit. We now need to increase the dimension $s$ of the
field beyond $s=2$. However, the problem of solving the unitarity equations
(\ref{eq:unitarity}) becomes increasingly difficult, since the unknown are matrices of increasingly
larger dimension $s\ge 3$ (we remind that the equations are bilinear non homogeneus in the unknown
transition matrices, and a canonical procedure for the solution is unknown). What we can do for
the moment is to provide only some particular solutions using algebraic techniques. Two ways of
obtaining solutions for $s=4$ is to start from solutions in dimension $s=2$ and built the direct-sum
and tensor product of two copies of the quantum walk in such a way that the obtained quantum walk
for dimension $s=4$ still satisfies the principles. We will see that the quantum walks that we
obtain in the relativistic limit give the Dirac equation when using the direct sum, whereas they give the
Maxwell equation (plus a static scalar field) when we use the tensor product.

When building a quantum walk in $2\times 2$ block form, all four blocks must be quantum walks
themselves. The requirement of locality of the coupling leads to off-diagonal blocks that do not
depend on $\bk$. A detailed analysis of the restrictions due to the unitarity conditions
(\ref{eq:unitarity}) shows that, modulo unitary change of representation independent on
$\bk$,\footnote{This can also be \eg the case of an overall phase independent of $\bk$.} we can take
the off-diagonal matrix elements as proportional to the identity, whereas the diagonal blocks are
just given by the chosen quantum walk and its adjoint, respectively. We then need to weight the
diagonal blocks with a constant $n$ and the off-diagonal identities with a constant $m$, and
unitarity requires having $|n|^2+|m|^2=1$.  Then, starting from the walk $A_\bk$ that leads us to the
Weyl equations for all dimension $d=1,2,3$, the walk, modulo unitary equivalence,\footnote{Also the
  solutions with walk $B^\pm=(A_\bk)^T$ are contained in Eq.  (\ref{eq:dirac-gen}), since they can
  be achieved either by a shift in the Brillouin zone or as $\sigma_yB^\pm\sigma_y=A^\pm{}^\dag$,
  with the exchange of the upper and lower diagonal blocks that can be done unitarily.} can be
recast in the form\cite{PhysRevA.90.062106}
\begin{equation}\label{eq:dirac-gen}
  D_\bk:=
  \begin{pmatrix}
    n A_\bk&im\\
    im&nA_\bk^\dag
  \end{pmatrix},\qquad n^2+m^2=1,\;n\in\Reals^+,m\in\Reals.
\end{equation}
Also the sign of $m$ can be changed by a unitary equivalence (a ``charge-conjugation''), however, we
keep $m$ with changing sign for reasons that will explained in Subsect. \ref{s:propertime}. The walk
(\ref{eq:dirac-gen}) with $s=4$ can be conveniently expressed in terms of gamma matrices in the spinorial
representation as follows
\begin{align}\label{eq:dirac}
  D_\bk:= nI u_\bk-i n\gamma^0\boldsymbol\gamma\cdot\tilde\bn_\bk+im\gamma^0,
\end{align}
where the functions $u_\bk$ and $\tilde\bn_\bk$ depend on the choice of $A_\bk$ in
Eq.~\eqref{eq:dirac-gen}, \ie on $d=1,2,3$. The dispersion relation of the quantum walk
\eqref{eq:dirac} is simply given by 
\begin{equation}
  \omega_\bk=\arccos[\sqrt{1-m^2}u_\bk].
\end{equation}
We will see now that the quantum walks in Eq.~\eqref{eq:dirac-gen} in the small wave-vector limit and
for $m\ll 1$ all give the usual Dirac equation in the respective dimension $d$, with $m$
corresponding to the particle rest mass, whereas  $n$ works as the inverse of a refraction index of
vacuum. In fact, the interpolating Hamiltonian $H(\bk)$ is given by
\begin{align}
  H(\bk)=\frac{\omega_{\bk}}{\sin\omega_{\bk}}(n\gamma^0\boldsymbol\gamma\cdot\tilde\bn_\bk-m\gamma^0), 
\end{align}
with relativistic limit given by
\begin{equation}\label{ref-index}
H_0(\bk)=\frac n{\sqrt d}\gamma^0\boldsymbol\gamma \cdot\bk+m\gamma^0,
\end{equation}
and to the order $\mathcal O(m^2)$ we get the Dirac Hamiltonian
\begin{equation}
H_0(\bk)=\frac 1{\sqrt d}\gamma^0\boldsymbol{\gamma}\cdot\bk+m\gamma^0.
\end{equation}
One has the Dirac Hamiltonian, with the wave-vector $\bk$ interpreted as momentum and the parameter
$m$ interpreted as the rest mass of the particle. In the relativistic limit (\ref{ref-index})
the parameter $n$ plays the role of the inverse of
a refraction index of vacuum. In principle this can produce measurable effects from bursts of  high-energy 
particles of different masses at the boundary of the visible universe, and would be complementary to the 
dispersive nature of vacuum (see subsections \ref{s:phydim} and \ref{s:vacdisp}).

\medskip

In the following we will also call the quantum walk in Eq. (\ref{eq:dirac-gen})  
{\em Dirac quantum walk}.\footnote{For $d=1$, modulo a permutation of the canonical basis, the
  quantum walk corresponds to two identical and decoupled $s=2$ walks. Each of these quantum
  walks coincide with the one dimensional {\em Dirac walks} derived in Ref.~\cite{Bisio2015244}.
  The last one was derived as the simplest $s=2$ homogeneous quantum cellular walk covariant
  with respect to the parity and the time-reversal transformation, which are less restrictive than
  isotropy that singles out the only Weyl quantum walk in one space dimension.}

In Ref. \cite{path1d} the discrete Feynman propagator for the Dirac quantum walk has been analytically 
evaluated with a closed formal  for dimension $d=1$, generalizing the solution of Ref. \cite{kauffman1996discrete} 
for fixed mass value.

\subsubsection{Discriminability between quantum walk and quantum field dynamics}
In Subsect. \ref{ss:limit} we mentioned that rigorous quantitative approaches to judge the closeness 
between the two dynamics have been provided in Ref.~\cite{Bisio2015244}, and in 
Ref.~\cite{PhysRevA.90.062106} in terms of fidelity between the two unitary evolutions. 
For the Dirac quantum walk for a proton mass one has fidelity close to unit for 
$N\simeq m^{-3}=2.2*10^{57}$, corresponding to $t=1.2*10^{14}\text{s}=3.7*10^6$  
years. The approximation is still good in the ultra-relativistic case $k\gg m$, \eg for $k=10^{-8}$
(as for an ultra-high energy cosmic ray), where it holds for $N\simeq k^{-2}=10^{16}$ steps, 
corresponding to $5*10^{-28}$ s. However, one should notice that practically the discriminability 
in terms of fidelity corresponds to having unbounded technology, and such a short time very likely 
corresponds to unfeasible experiments. On the other hand, for a ultra-high energy proton 
with wave packet width of  100 fm the time required for discriminating the wave-packet  of the
quantum walk from that of QFT is comparable with the age of the universe.

\subsubsection{Mass and proper-time}\label{s:propertime}
The unitarity requirement in Eq. (\ref{eq:dirac-gen}) restrict the rest mass to belong to the interval 
\begin{equation}
m\in[-1,1].
\end{equation}
At the extreme points $\pm 1$ of the interval the corresponding dynamics $D_\bk=\pm i\gamma^0$ are
identical (they differ for an irrelevant global phase factor). This means that the domain of the
mass has actually the topology of a circle, namely
\begin{equation}
m\in S^1.
\end{equation}
From the classical relativistic Hamiltonian \cite{greenberger1970theory}
\begin{equation}
H=\vec p\cdot\vec q+c^2 m\tau-L,
\end{equation}
with $\vec p$ and $\vec q$ canonically conjugated position and momentum and $L$ the Lagrangian, we
see that the {\em proper time} $\tau$ is canonically conjugated to the rest mass $m$. This suggests
that the Fourier conjugate of the rest mass in the quantum walk can be interpreted as the proper
time of a particle evolution, and being the mass a variable in $S^1$, we conclude that {\em the
  proper time is discrete}, in accordance with the discreteness of the dynamical evolution of the
quantum walk. This result constitutes a non trivial logical coherence check of the present quantum 
walk theory.

\subsubsection{Physical dimensions and scales for mass and discreteness}\label{s:phydim}
We want to emphasize that in the above derivation everything is adimensional by construction.
Dimensions can be recovered by using as measurement standards for space, time, and mass the
discreteness scale for space $a_*$ and time $t_*$ ($a_*$ is half of the BCC cell side, $t_*$ the
time-length of the unit step), along with the maximum value of the mass $m_*$ (corresponding to
$|m|=1$ in Eq. \eqref{eq:dirac-gen}).  From the relativistic limit, the comparison with the usual
dimensional Dirac equation leads to the identities
\begin{equation}\label{Planck}
c=a_*/t_*,\quad \hbar=m_*a_*c,
\end{equation}
which leave only one unknown among the three variables $a_*,t_*$ and $m_*$.  At the maximum value of
the mass $|m|=1$ in Eq.  \eqref{eq:dirac-gen} we get a flat dispersion relation, corresponding to no
flow of information: this is naturally interpreted as a mini black-hole, \ie a particle with Schwarzild
radius equal to the localization length, \ie the Compton wavelength. This leads to an
heuristic interpretation of $m_*$ as the Planck mass, and from the two identities in Eq.
(\ref{Planck}) we get the Planck scale for discreteness. Notice that the value of $m_*$ can be in
principle obtained from the dispersion of vacuum as $m_*\simeq\frac{1}{\sqrt{3}}\frac{\hbar
  k}{c(k)-c(0)}$ for small $k$, which can be in principle measured by the Fermi telescope from
detection of ultra high energy bursts coming from deep space.

\subsection{Recovering Maxwell fields\protect\footnote{The entire subsection is a short
    review of Ref.  \cite{Bisio2016}}}\label{s:maxwell}

In Sections \ref{s:weyl} and \ref{s:dirac} we showed how the dynamics of free quantum fields can
be derived starting from a countable set of quantum systems with a network of interactions
satisfying the principles of locality, homogeneity, and isotropy. Within the present finitistic
local-algorithmic perspective one also considers each system as carrying a finite amount of
information, thus restricting the quantum field to be Fermionic (see also Subsect.
\ref{s:bosonferm}). However, one may wonder how the physics of the free electromagnetic field can be
recovered in such a way and, generally, how Bosonic fields are recovered from Fermionic ones.  In
this section we answers to these questions.  The basic idea behind is that {\em the
photon emerges as an entangled pair of Fermions evolving according to the Weyl quantum walk} of
Section \ref{s:weyl}. Then one shows that in a suitable regime
both the free Maxwell equation in 3d and the Bosonic commutation relations are recovered. Since in this subsection 
we are actually considering operator quantum fields, we will use more properly the quantum automaton nomenclature 
instead of the quantum walk one.

Consider two Fermionic fields $\psi(\bk)$ and $\varphi(\bk)$ in the wave-vector representation, with
respective evolutions given by
\begin{align}
  \label{eq:automa2}
  {\psi} (\bk,t+1) = W_\bk{\psi} (\bk,t).
\quad
  {\varphi} (\bk,t+1) = W_\bk^*{\varphi} (\bk,t).
\end{align}
The matrix $W_\bk$ can be any of the Weyl quantum walks for $d=3$ in Eq.~\eqref{eq:weyl3d}, (the
whole derivation is independent on this choice), whereas $W_\bk^* = \sigma_y W_\bk\sigma_y$ denotes
the complex conjugate matrix. We introduce the bilinear operators
\begin{equation}
G^{i}(\bk,t) :=\varphi^T (\tfrac{\bk}{2},t) \sigma^{i}  \psi(\tfrac{\bk}{2} , t)=\varphi^T (\bk,0) (
W_{\tfrac{\bk}{2}}^\dag  \sigma^{i} W_{\tfrac{\bk}{2}} )\psi(\tfrac{\bk}{2} , 0) 
\end{equation}
by which we construct the vector field
\begin{equation}
\bvec{G}(\bk,t):= ( G^{1}(\bk,t), G^{2}(\bk,t), G^{3}(\bk,t))^T
\end{equation}
and the transverse field
\begin{equation} \label{eq:prephoton}
 \bvec{G}_T(\bk,t):=  \bvec{G}(\bk,t) -  
\left(\frac{\bvec{n}_{\frac{\bk}{2}}}{|\bvec{n}_{\frac{\bk}{2}}|} \cdot
{\bvec{G}}(\bk,t) \right) \frac{\bvec{n}_{\frac{\bk}{2}}}{|\bvec{n}_{\frac{\bk}{2}}|},
\end{equation}
with $\bn_\bk:= (\omega_{\bk}/\sin\omega_{\bk})\tilde{\bn}_{\bk}$ and $\tilde\bn_\bk$ given in
Eq. (\ref{eq:weyl3dxyz}). By construction the field $\bvec{G}_T(\bk,t)$ satisfies 
the following relations
\begin{align}
  \label{eq:premaxwell}
 \bvec{n}_{\frac{\bk}{2}} \cdot \bvec{G}_T(\bk,t)  &=  0,\\
  \bvec{G}_T(\bk,t) &= \Exp(-i2\bvec{n}_{\tfrac{\bk}{2}} \cdot \bvec{J}
  t)  \bvec{G}_T(\bk,0), \label{eq:premaxwell2}
\end{align}
where we used the identity 
\begin{equation}
 \exp (-\tfrac{i}{2}\bvec{v}\cdot \boldsymbol{\sigma}) 
\boldsymbol{\sigma} \exp (\tfrac{i}{2}\bvec{v} \cdot \boldsymbol{\sigma}) =
\Exp(-i\bvec{v} \cdot \bvec{J}) \boldsymbol{\sigma},
\end{equation}
the matrix $\Exp(-i\bvec{v}\cdot\bvec{J})$ acting on $\boldsymbol{\sigma}$ regarded as a vector,
and $\bvec J=(J_x, J_y,J_z)$ representing the infinitesimal generators of $\SU(2)$ in the spin 1 representation.
Taking the time derivative of Eq. \eqref{eq:premaxwell2} we obtain
\begin{align}
  \label{eq:premaxwell3}
  \partial_t\bvec{G}_T(\bk,t) = 2\bvec{n}_{\tfrac{\bk}{2}} \times  \bvec{G}_T(\bk,t).
\end{align}
If $\bvec{E}_G$ and
$\bvec{B}_G$ are two Hermitian operators defined by the relation
\begin{align}
  \label{eq:electric and magnetic field}
  \bvec{E}_G:=|{\bn}_{\tfrac\bk2}|(\bvec{G}_T+\bvec{G}_T^\dag),\quad\bvec{B}_G:=i|{\bn}_{\tfrac\bk2}|(\bvec{G}_T^\dag-\bvec{G}_T),
\end{align}
then Eq. \eqref{eq:premaxwell} and Eq. \eqref{eq:premaxwell3}  
can be rewritten as 
\begin{align}
   & \partial_t \bvec{E}_G = i 2\bvec{n}_{\tfrac{\bk}{2}} \times
    \bvec{B}_T(\bk,t)  
&&\partial_t \bvec{B}_G =- i 2\bvec{n}_{\tfrac{\bk}{2}} \times \bvec{E}_T(\bk,t)    \nonumber\\
          & 2\bvec{n}_{\tfrac{\bk}{2}} \cdot \bvec{E}_G = 0 
&& 2\bvec{n}_{\tfrac{\bk}{2}} \cdot \bvec{B}_G = 0. 
      \label{eq:maxweldistorted}
 \end{align}
 Eqs. (\ref{eq:maxweldistorted}) have the form of distorted Maxwell equations, with the wave-vector
 $\bk$ substituted by $ 2\bvec{n}_{\tfrac{\bk}{2}}$, and in the relativistic limit $|\bk| \ll 1$ one
 has $ 2\bvec{n}_{\tfrac{\bk}{2}} \sim \bk$ and the usual free electrodynamics is recovered.

\subsubsection{Photons made of pairs of Fermions}
Since in the Weyl equation the field is Fermionic, the field defined in Eqs. \eqref{eq:prephoton}
and \eqref{eq:electric and magnetic field} generally does not satisfy the correct Bosonic
commutation relations.  The solution to this problem is to replace the operator $\bvec{G}$ defined
in Eq.  \eqref{eq:prephoton} with the operator $\bvec{F}$ defined as
\begin{align}
  \label{eq:photon}
\bvec{F}(\bk) := 
 \int \frac{ d \bvec{q}}{(2 \pi)^3}
f_{\bk}(\bvec{q})
\varphi
\left(\tfrac{\bk}{2}-\bvec{q}\right)
\,\boldsymbol{\sigma}\,
\psi
 \left(\tfrac{\bk}{2}+\bvec{q}\right), 
\end{align}
where
$\int\frac{d\bvec{q}}{(2\pi)^3} |f_{\bk}(\bvec{q})|^2 =1, \forall \bk$.
In terms of $\bvec{F}(\bk)$, we can define the polarization operators 
$\varepsilon^i(\bk)$
of the electromagnetic field as follows
\begin{align}
 &\varepsilon^i(\bk) := \bvec{u}^i_\bk\cdot\bvec{F}(\bk,0),\quad i=1,2,
  \label{eq:polarization}
\\
&\bvec{u}^i_\bk \cdot \bn_{\bk} =\bvec u^1_\bk\cdot\bvec u^2_\bk= 0,
\;
 |\bvec u^i_\bk|=1,
\;
(\bvec u^1_\bk\times\bvec u^2_\bk)\cdot\bn_\bk>0.
\end{align}
In order to avoid technicalities from continuum of wavevectors, we restrict to a discrete
wave-vector space, corresponding to confinement in a cavity.  Moreover we assume $|
{f}_\bk(\bvec{q})|^2$ to be uniform over a region $\Omega_\bk$ which contains $N_\bk$ modes, \ie
\begin{equation}
|{f}_\bk(\bvec{q})|^2 =
\begin{cases}
\tfrac{1}{N_\bk} &\text{if }\bvec{q}\in \Omega_\bk\\
0&\text{ otherwise}.
\end{cases}
\end{equation}
Then, for a given state $\rho$ of the field we denote by $M_{\varphi,\bvec{k}}$ (resp.
$M_{\psi,\bvec{k}}$) the mean number of type $\varphi$ (resp $\psi$) Fermionic excitations in the
region $\Omega_\bk$.  One can then show that, for states such that $M_{\xi,\bvec{k}}/ N_\bk \leq
\epsilon\ll 1$ for both $\xi = \varphi, \psi$ and forall $\bvec{k}$ we have 
\begin{equation}
[\varepsilon^i (\bvec{k}),{\varepsilon^j}^\dag (\bvec{k}')]_- = \delta_{i,j} \delta_{\bvec{k},\bvec{k}'},
\end{equation}
\ie the polarization operators are Bosonic operators.

\begin{figure}[t]
  \begin{center}
    \includegraphics[width=.49\textwidth]{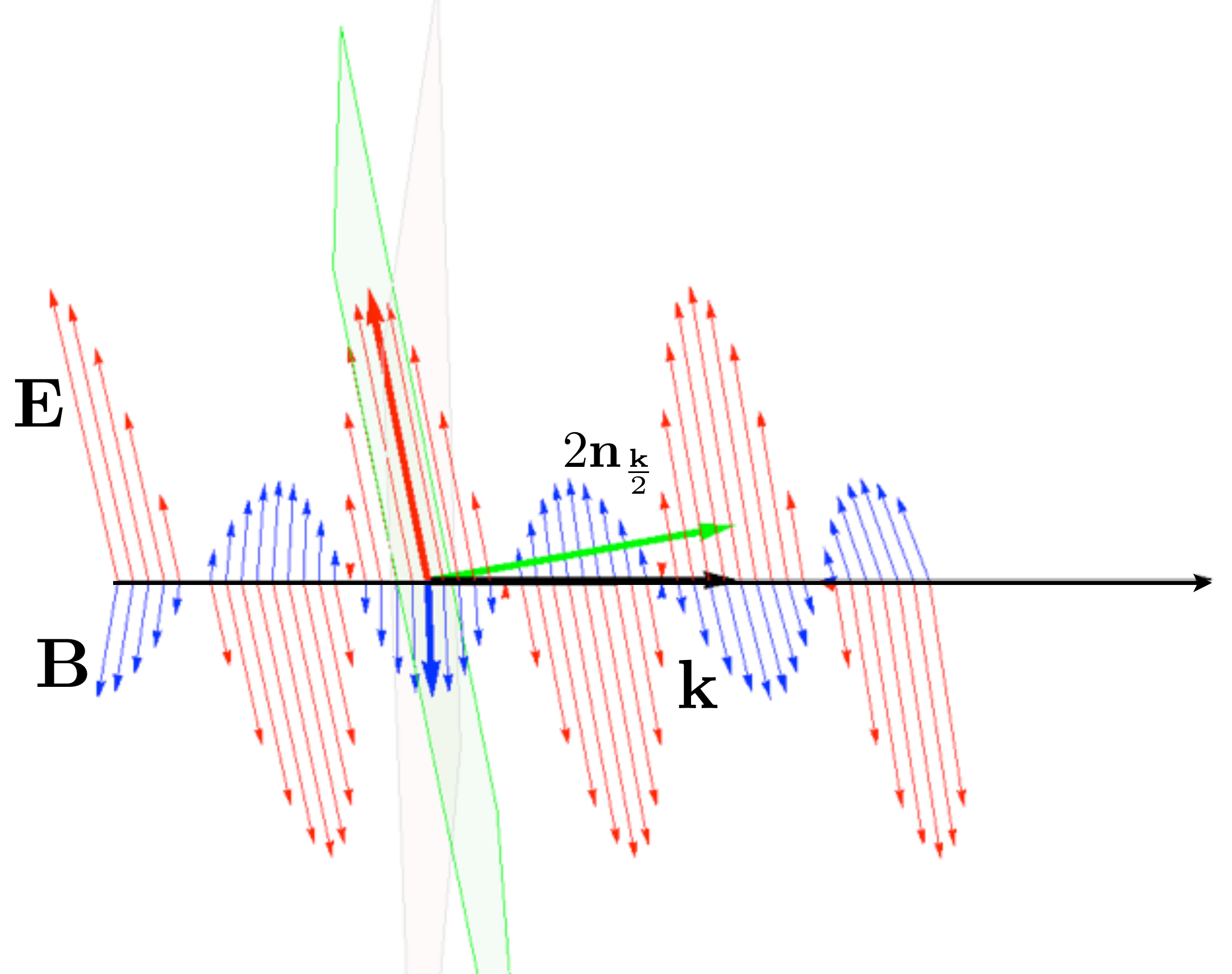}
    \includegraphics[width=.49\textwidth]{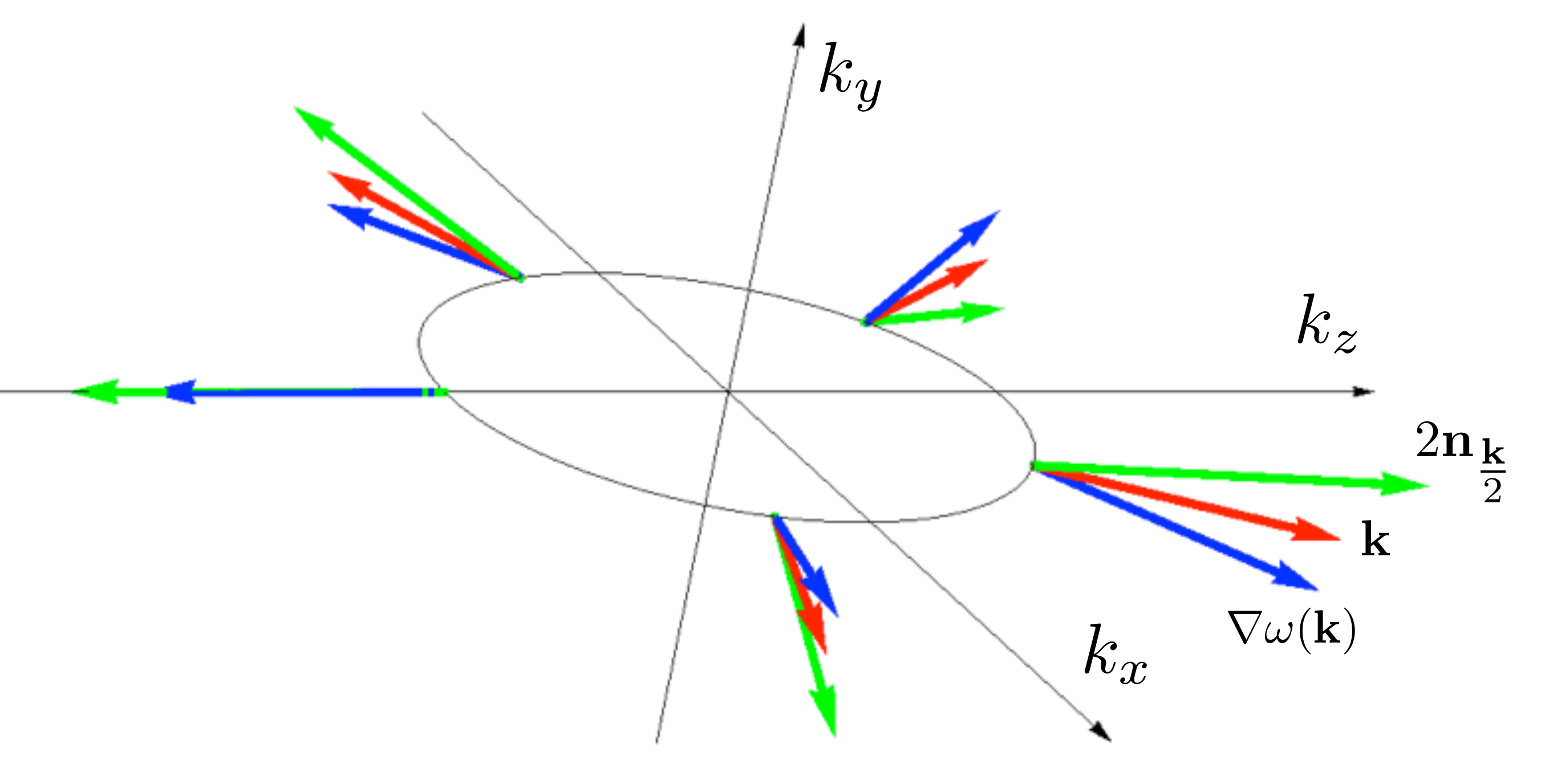}
    \caption{From Ref.~\cite{Bisio2016}. (Colors online). Left: In a rectilinear polarized
      electromagnetic wave, the polarization plane (in green) is slightly tilted with 
      respect the plane orthogonal to $\bk$ (in gray). Right: vector $2\bvec{n}_{\tfrac{\bk}{2}}$
      (in green), which is orthogonal to the polarization plane; wavevector $\bk$ (in red) and group
      velocity (in blue) for the value $|\bk|= 0.8$ and different directions. Notice that the three
      vectors are not parallel (the angles between them depend on $\bk$).\label{fig:relvectors}}
  \end{center}
\end{figure}
\subsubsection{Vacuum dispersion \label{s:vacdisp}}
According to Eq. \eqref{eq:maxweldistorted} the angular frequency of the electromagnetic waves is
given by the modified dispersion relation 
\begin{align}
\label{eq:modifieddisprelmax}
\omega(\bk) = 2 | \bvec{n}_{\tfrac{\bk}{2}} |,
\end{align}
which recovers the usual relation $\omega(\bk) = | \bk | $ in the relativistic regime. In a
dispersive medium, the speed of light is the group velocity $\nabla_\bk \omega(\bk)$ of the
electromagnetic waves, and Eq. \eqref{eq:modifieddisprelmax} predict that the vacuum is dispersive,
namely the speed of light generally depends on $\bk$. Such dispersion phenomenon has been already
analyzed in some literature on quantum gravity, where several authors considered how an hypothetical
invariant length (corresponding to the Planck scale) could manifest itself in terms of modified
dispersion relations
\cite{ellis1992string,lukierski1995classical,Quantidischooft1996,amelino2001testable,PhysRevLett.88.190403}.
In these models the $\bk$-dependent speed of light $c(\bk)$, at the leading order in $k :=| \bk |$,
is expanded as $c(\bk) \approx 1 \pm \xi k^{\alpha}$, where $\xi $ is a numerical factor of the
order $1$, while $\alpha$ is an integer.  This is exactly what happens in our framework, where the
intrinsic discreteness of the quantum cellular automata $A_\bk^\pm$ leads to the dispersion relation of
Eq. \eqref{eq:modifieddisprelmax} from which one obtains the following $\bk$-dependent speed of
light
\begin{align} \label{eq:freqdepsol}
  c^\mp(\bk) \approx 1 \pm 3\frac{k_x k_y k_z}{|\bk|^2} \approx
 1 \pm \tfrac{1}{\sqrt{3}}k.
\end{align}
Eq. (\ref{eq:freqdepsol}) is obtained by evaluating the modulus of the group velocity and expanding
in powers of $\bk$ with the assumption $ k_x = k_y = k_z = \tfrac{1}{\sqrt{3}} k $, ($k =
|\bk|$).\footnote{Notice that, depending on the quantum walk $A^{+}(\bk)$ of $A^{-}(\bk)$ in Eq.
  (\ref{eq:weyl3d}) we obtain corrections to the speed of light with opposite sign.} Notice that the
dispersion is not isotropic, and can also be superluminal, though uniformly bounded
\cite{PhysRevA.90.062106} by a factor $\sqrt{d}$ (which coincides with the uniform bound of the
quasi-isometric embedding). The prediction of dispersive behavior, as for the present automata
theory of quantum fields, is especially interesting since it is experimentally falsifiable, and, as
mentioned in Subsect. \ref{s:phydim}, allows to experimentally set the discreteness scale. In fact,
differently to other birefringence effects (Fig.\ref{fig:relvectors}), the disperision effect, although is
extremely small in the relativistic regime, it accumulates and become magnified during a huge time
of flight. For example, observations of the arrival times of pulses originated at cosmological
distances (such as in some $\gamma$-ray
bursts\cite{amelino1998tests,abdo2009limit,vasileiou2013constraints,amelino2009prospects}), have
sufficient sensitivity to detect corrections to the relativistic dispersion relation of the same
order as in Eq. \eqref{eq:freqdepsol}.

\section{Recovering special relativity in a discrete quantum 
universe\protect\footnote{This entire section is a review of the main results of
  Ref. \cite{lrntz3d}.}}\label{s:SR}
We have seen how relativistic mechanics, and more precisely free QFT, can be recovered without using
any mechanical primitive, and without making any use of special relativity, including the relativity
principle itself. However, one may wonder how discreteness can be reconciled with Lorentz
transformations, and most importantly, how the relativity principle itself can be restated in purely
mathematical terms, without using the notions of space-time and inertial frame. In this section 
we will see how such goal can be easily accomplished.

\medskip
The relativity principle is expressed by the statement:
\begin{enumerate}
\item[] {\em Galileo's Relativity Principle:} The physical law is invariant with the inertial
  frame.
\end{enumerate}
Otherwise stated: the physics that we observe, or, equivalently, its mathematical representation, is independent on the inertial frame that we use. 

What is a frame?  It is a mathematical representation of physical laws in terms of space and time coordinates. What is
special about the {\em inertial} frame? A convenient way of answering is the following 
\begin{enumerate}
\item[] {\em Inertial frame:} a reference frame where energy and momentum are conserved for
  an isolated system.
\end{enumerate}
When a system is isolated? This is established by the theory. In classical mechanics, a system is
isolated if there are no external forces acting on it. In quantum theory a system is isolated when its
dynamical evolution is described by a unitary transformation on the system's Hilbert space. At the
very bottom of  its notion, the inertial frame is the mathematical representation of the physical law
that makes its analytical form the simplest. In classical physics, if we include the Maxwell equations among the invariant
physical laws, what we get from Galileo's principle is Einstein's special relativity.

The quantum walk/automaton is an isolated system (it evolves unitarily). Mathematically the
physical law that brings the information about the constants of the dynamics in terms of their
Hilbert eigenspaces is provided by the eigenvalue equation. For the case of virtually Abelian group
$G$ (which ultimately leads to physics in Euclidean space) the eigenvalue equation has the general
form corresponding to Eqs.  (\ref{eq:diff}) and (\ref{e:omega})
\begin{equation}\label{s:eigA}
A_{\v{k}}\psi(\omega,\v{k})=e^{i\omega}\psi(\omega,\v{k}),
\end{equation}
with the eigenvalues usually collected into $s$ dispersion relations (the two functions
$\omega^\pm(\bk)$ for the Weyl quantum walk). This translates into the following re-interpretation of
representations of the eigenvalue equation:
\begin{enumerate}
\item[] {\em Quantum-digital inertial frame:} Representation in terms of eigen-spaces of the
  constants of the dynamics of the eigenvalue equation (\ref{s:eigA}).
\end{enumerate}
Using such notion of inertial frame, the principle of relativity is still the Galileo's principle.  The
group of transformations that connect different inertial reference frames will be the quantum
digital-version of the Poincar\'e group:
\begin{enumerate}
\item[] {\em Quantum-digital Poincar\'e group:} group of changes of representations in terms
  of eigenspaces of the dynamical constants that leave the eigenvalue equation (\ref{s:eigA}) invariant.
\end{enumerate}
It is obvious that the changes of representations make a group. Since the constants of dynamics are $\bk$ and $\omega^\pm$, a change of representation corresponds to an invertible map  $k\to k'(k)$, where with $k$ we denote the four-vector $k:=(\omega,\bk)$. 

In the following subsection we will see how the inherent discreteness of the algorithmic 
description leads to distortions of the Lorentz transformations, visible in principle 
at huge energies. Nevertheless, Einstein's special relativity is perfectly recovered for $|\bk|\ll 1$, 
namely at energy scales much higher than those ever tested. 

On the other hand, as we will see in the following, discreteness has some plus compared to the continuum theory, 
since it contains the continuum theory as a special regime, and moreover it leads to some additional features with 
GR flavor: 1) it has a maximal particle mass with physical interpretation in terms of the Planck mass; 2) it leads to a De Sitter 
invariance (see Subsect. \ref{s:DeSitt}). And this, in addition to providing  its own physical standards for space, 
time, and mass within a purely mathematical context (Subsect. \ref{s:phydim}). 

\subsection{Quantum-digital Poincar\'e group and the notion of particle\protect\footnote{For a simpler analysis in one space dimensions and the connection with doubly-special relativity and relative locality, see Ref.\cite{bibeau2013doubly}. For a connection with Hopf algebras for position and momentum see Ref. \cite{hopf}.}}
The eigenvalue equation (\ref{s:eigA}) can now be rewritten in ``relativistic notation'' as follows
\begin{align}
\label{eq:hamiltonian2}
n_\mu(k)\sigma^\mu\psi(k) = 0,
\end{align}
upon introducing the four-vectors
\begin{equation}
k=(\omega,\bk),\quad n(k)=(\sin\omega,\v{n}(\bk)),\quad
\boldsymbol{\sigma}=(I,\v{\sigma}),\quad \v{\sigma}=(\sigma_x,\sigma_y,\sigma_z),
\end{equation}
where the vector $\v{n}(\bk)$ is defined in Eq. (\ref{eq:weyl3d}), namely 
\begin{equation}
\label{eq:generalautomaton}
\v{n}(\v{k})\cdot\v{\sigma}: = \frac{i}{2}(A_{\v{k}} -A_{\v{k}}^\dagger).
\end{equation}
As already mentioned, since the constants of dynamics are $\bk$ and $\omega^\pm$, a change of representation corresponds
 to a map $k\mapsto k'(k)$. Now the principle of relativity corresponds to the requirement that the
 eigenvalue equation (\ref{eq:hamiltonian2}) is preserved under a change of representation. This
 means  that the following identity must hold
\begin{align}
\label{eq:invariantdynam}
n_\mu(k)\sigma^\mu=
\tilde\Gamma^{-1}_k\,n_\mu(k')\sigma^\mu\,\Gamma_k,
\end{align}
where $\Gamma_k$, $\tilde{\Gamma}_k$ are invertible matrices representing the change of
representation. 
 \begin{figure*}[t]
\begin{center}
 \includegraphics[width=\textwidth]{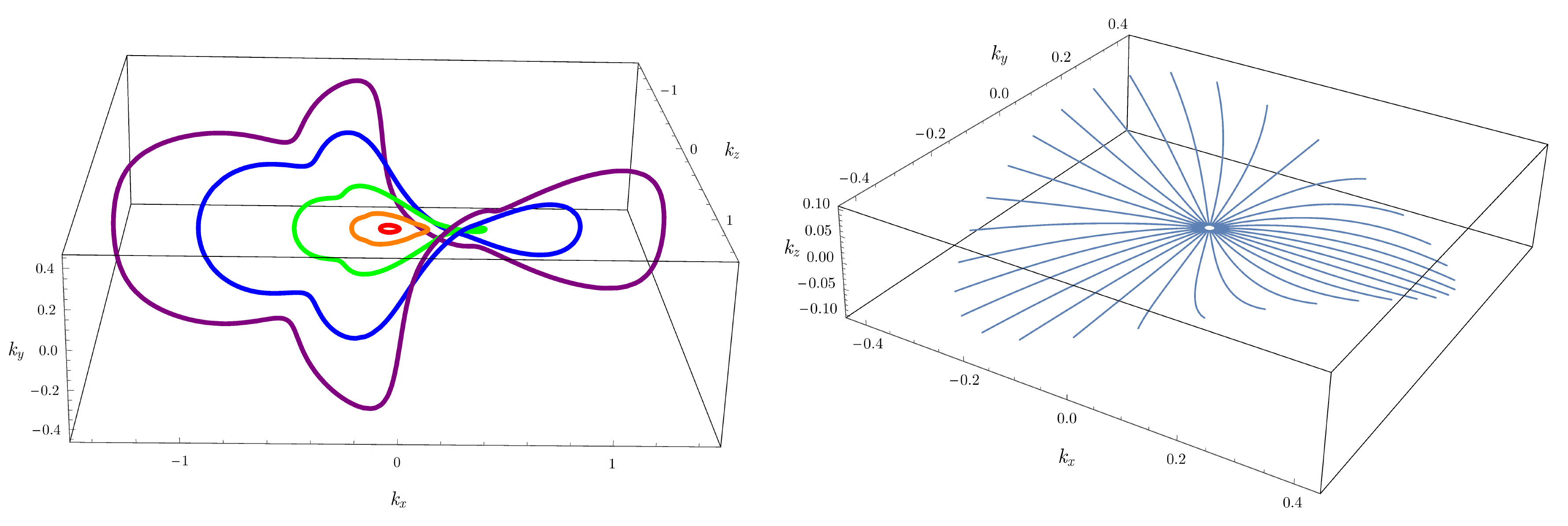}\\
 \includegraphics[width=.5\textwidth]{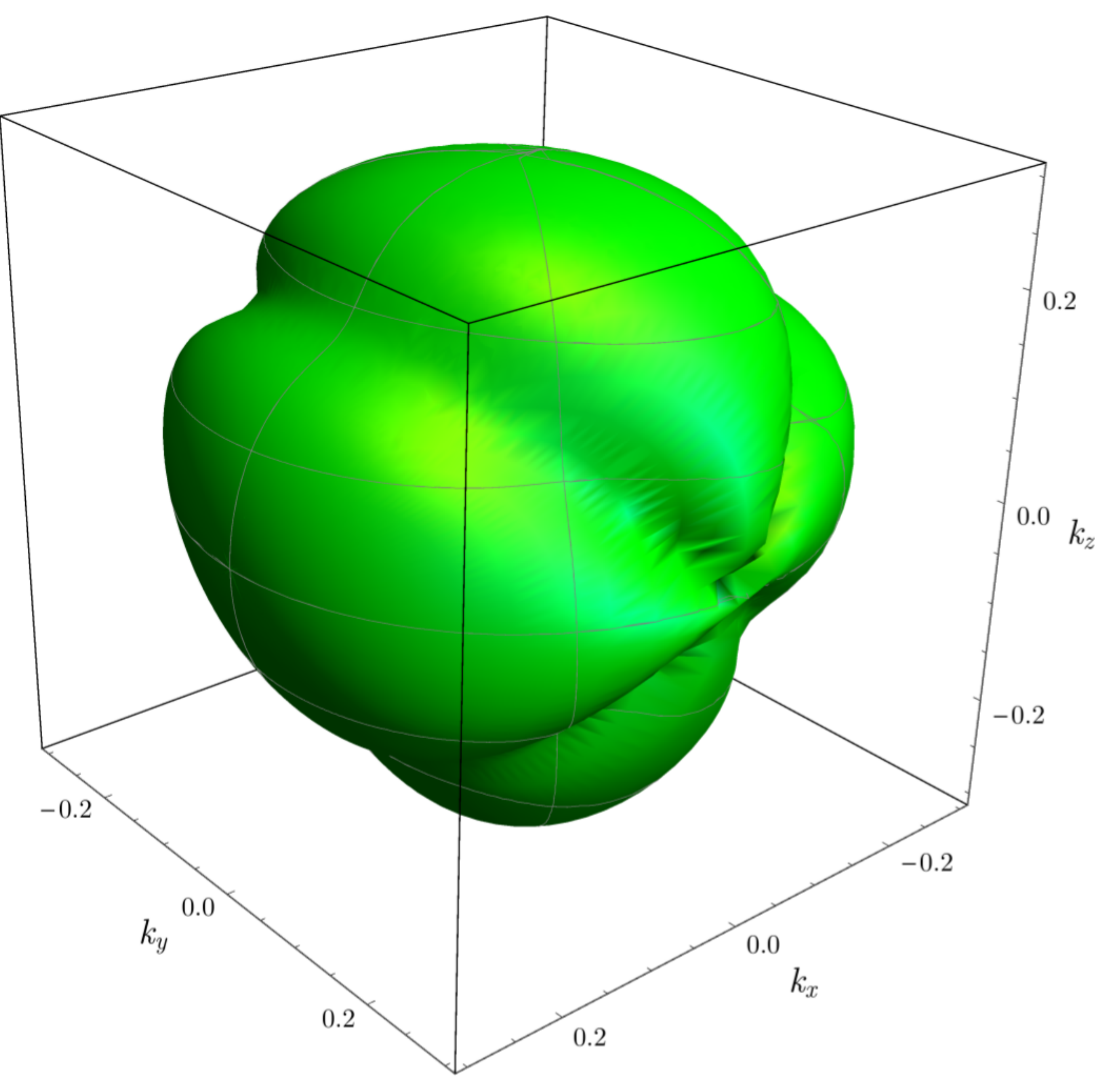}
\end{center}
 \caption{From Ref. \cite{lrntz3d}. (Colors online). The distortion effects of the Lorentz group in the present quantum walk
   theory leading to the Weyl quantum field in the relativistic limit. 
  {\em Top left figure}: the orbit of the wavevectors $\bk=(k_x,0,0)$, with
  $k_x\in\{.05,.2,.5,1,1.7\}$ under the rotation around the $z$ axis.  
 {\em Top right figure:} the orbit of wavevectors with $|\bk|=0.01$ for various directions in the
 $(k_x,k_y)$ plane under the boosts with $\bbeta$ parallel to $\bk$ and $|\bbeta|\in[0,\tanh
 4]$. {\em Bottom figure:} the orbit of the wavevector $\bk=(0.3,0,0)$ under the full rotation group $SO(3)$.\label{orb2d}}
 \end{figure*}

The simplest example of change of observer is the one given by the trivial relabeling $k'=k$ and by
the matrices $ \Gamma_k=\tilde{\Gamma}_k=e^{i\lambda(\v{k})}$, where $\lambda(\v{k})$ is an
arbitrary real function of $\v{k}$.  When $\lambda(\v{k})$ is a linear function we recover the usual
group of translations. The set of changes of representation $k\mapsto k'(k)$ for which Eq.
(\ref{eq:invariantdynam}) holds are a group, which is the largest group of symmetries of the
dynamics. In covariant notation the dispersion relations are rewritten as follows
\begin{align}
\label{eq:eigenequation2}n_\mu^\pm(k)n^{\mu\pm}(k)=0,  
\end{align}
and in the small wave-vector regime one has $n(k)\sim k$, recovering the usual relativistic
dispersion relation. 

In addition to the neighbour of the wavevector $\v{k_0}=(0,0,0)$, the Weyl equations can be
recovered from the quantum walk (\ref{eq:weyl3d}) also in the neighborhood of the wavevectors in Eq.
(\ref{e:kkk}). The mapping between the vectors $\v{k}_i$ exchange chirality of the particle and
double the particles to four species in total: two left-handed and two right-handed.\footnote{Discreteness 
has doubled the particles: this corresponds to the well known phenomenon of Fermion
doubling \cite{PhysRevD.16.3031}.} In the following we will therefore more generally refer to the
relativistic regime as the neighborhoods of the vectors $\{\v{k}_i\}_{i=0}^3$.

The group of symmetries of the dynamics of the quantum walks (\ref{eq:weyl3d}) contains a nonlinear
representation of the Poincar\'e group, which exactly recovers the usual linear one in the
relativistic regime.  For any arbitrary non vanishing function $f(k)$ one introduces the four-vector
\begin{align}
\label{eq:fff}
p^{(f)}=\mathcal{D}^{(f)}(k):=f(k)n(k) 
\end{align}
and rewrite the eigenvalue equation (\ref{eq:hamiltonian2}) as follows 
\begin{align}
  p^{(f)}_\mu \sigma^{\mu} \psi(k) = 0.
\end{align}
Upon denoting the usual Lorentz transformation by $L_\bbeta$ for a suitable $f$ \cite{lrntz3d} the Brillouin zone
splits into four regions $\Brill_i$, $i=1,\ldots,4$ centered around $\v{k}_i$ $i=0,\ldots 3$, such
that the composition
\begin{align}\label{calL}
  \mathcal{L}^{(f)}_\bbeta := \mathcal{D}^{(f)-1} L_\bbeta \mathcal{D}^{(f)}
\end{align}
is well defined on each region separately. The four invariant regions corresponding to the four
different massless Fermionic particles show that the Wigner notion of ''particle'' as invariant of
the Poincar\'e group survives in a discrete world. For fixed function $f$ the maps
$\mathcal{L}^{(f)}_\bbeta$ provide a non-linear representation of the Lorentz group
\cite{amelino2001planck,amelino2002relativity,magueijo2002lorentz}.  In Figs. \ref{orb2d} the orbits
of some wavevectors under subgroups of the nonlinear Lorentz group are reported. The distortion
effects due to underlying discreteness are evident at large wavevectors and boosts. The relabeling
$k\rightarrow k'(k)=\mathcal{L}^{(f)}_\bbeta(k)$ satisfies Eq. (\ref{eq:invariantdynam}) with
$\Gamma_k=\Lambda_\bbeta$ and $\tilde\Gamma_k=\tilde\Lambda_\bbeta$ for the right-handed particles,
and $\Gamma_k=\tilde\Lambda_\bbeta$ and $\tilde\Gamma_k=\Lambda_\bbeta$ for the left-handed
particles, with $\Lambda_\bbeta$ and $\tilde\Lambda_\bbeta$ being the $(0,\tfrac12)$ and
$(\tfrac12,0)$ representation of the Lorentz group, independently on $k$ in each pertaining region.

For varying $f$, one obtains a much larger group, including infinitely many copies of the nonlinear
Lorentz one. In the small wave-vector regime the whole group collapses to the usual linear Lorentz
group for each particle.

\subsection{De Sitter group for nonvanishing  mass}\label{s:DeSitt}
Up to now we have analyzed what happens with massless particles. For massive particles described by
the Dirac walk (\ref{eq:dirac-gen}), the rest-mass $m$ gets involved into the frame transformations, 
and their group becomes a nonlinear realization of the De Sitter group $SO(1,4)$
with infinite cosmological constant, where the rest mass $m$ of the particle plays the role of the
additional coordinate. One recovers the previous nonlinear Lorentz group at the order $\mathcal O(m^2)$.

\section{Conclusions and future perspectives: the interacting theory, ..., gravity?}\label{s:future}

\begin{figure}
\includegraphics[width=1\textwidth]{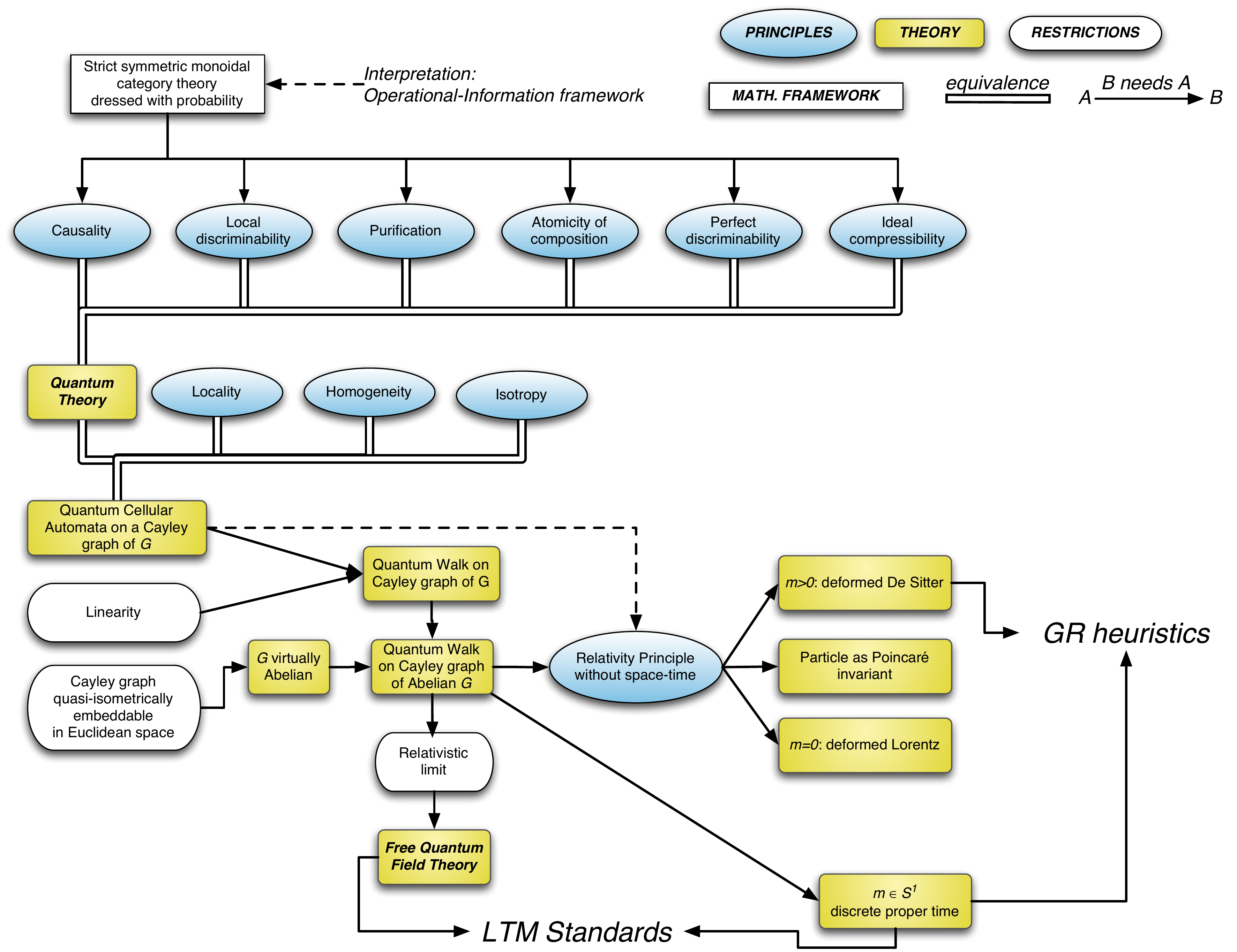}\\
\caption{Logical scheme of the derivation from principles of the present quantum-walk theory of fields, with the known free quantum field theory as its relativistic limit. The top six principles from which quantum theory of abstract system is derived  are not discussed in the present paper, and can be found in Refs. \cite{chiribella2011informational,QTfromprinciples}.\label{fig:logscheme}}
\end{figure}

The logical connections that have lead us to build up our quantum-walk theory 
of fields leading to free QFT are summarized in Fig. \ref{fig:logscheme}.
The free relativistic quantum field theory emerges as a special regime (the relativistic regime) of the evolution of 
countably many Fermionic quantum bits, provided that their unitary interactions satisfy the principles of 
homogeneity, locality, and isotropy, and with the restrictions of linearity of the evolution and of quasi-isometric 
embedding of the graph of interaction in an Euclidean space. 

We are left now with the not easy task of recovering also the interacting relativistic quantum field theory, where 
particles are created and annihilated. We will need to devise which additional principles are missing that will lead 
to the interacting theory, breaking the linearity assumption. This is likely to be related to the nature of  a gauge 
transformation. How can this be restated in terms of a new principle? From the point of view of
a free theory, the interaction can be viewed as a violation of homogeneity, corresponding to the presence of another 
interacting field--namely the gauge-field. The gauge-field  can be regarded as a restoration of homogeneity
by a higher level homogeneous ``meta-law''. For example, one can exploit the arbitrariness of the local bases of the Hilbert 
block subspaces $\Cmplx^s$ for the Weyl automata, having the bases dependent on the local value of the wave-function of 
the gauge automaton made with pairs of entangled Fermions, as for the Maxwell automaton. 
In order to keep the interaction local, one can consider an 
on-situ interaction. In such a way one would have a {\em quantum ab initio} gauge theory, without the need of 
artificially quantizing the gauge fields, nor of introducing mechanical Lagrangians. A $d=1$ interacting theory of the kind of a Fermionic Hubbard quantum cellular automaton, has been very recently analytically investigated by the Bethe ansatz \cite{Hubbard}, and two-particle bound states have been established. It should be emphasized that for $d=3$ 
just the possibility of recovering QED in the relativistic regime would be very interesting, since it will
provide a definite procedure for renormalization. Very interesting will be also the analysis of the full dynamical invariance 
group, leading also to a nonlinear version of the Poincar\'e group, with the possibility that this restricts the choice of the 
function $f$ in Eq. (\ref{eq:fff}). Studying the full symmetry group of the interacting theory will also have the potential of providing 
additional internal symmetries, \eg the $SU(3)$ symmetry group of  QCD, with the Fermion doubling possibly playing the role
in adding physical particles. The mass as a variable quantum observable (as in Subsections 
\ref{s:propertime} and \ref{s:DeSitt}) may provide rules about the lifetimes of different species of particles. 
The additional quasi-static scalar mode entering in the tensor-product of the two Weyl automata that give the Maxwell 
field in Subsect. \ref{s:maxwell} may turn out to play a role in the interacting theory, \eg playing the role of a Higg boson providing the 
mass value, or even being pivotal for gravitation. But for now we are just in the realm of speculations.

What we can say for sure is that it is not just a coincidence that so much physics comes out from so few general principles. 
How amazing is the whole resulting theory which, in addition to having a complete logical coherence by construction, it also winks 
at GR through the two nontrivial features of the maximum mass, and the De Sitter invariance. And 
with special relativity derived without space-time and kinematics, in a fully {\em quantum ab initio} theory. So much from 
so little? This is the power of the new information-theoretical paradigm.
 
\section*{Appendices}
\appendix
\section{A very brief historical account}\label{sec:his}
The first very preliminary heuristic ideas about the current quantum cellular automaton/quantum walk theory have been presented in a friendly and open-minded environment  at Perimeter Institute in Waterloo in a series of three talks in 2010-11 
\cite{darianopirsa-GUT1,darianopirsa-GUT2,darianopirsa-GUT3}.\footnote{Other talks have been presented in the V\"axj\"o conference on quantum foundations \cite{darianovaxjo2010,darianovaxjo2011,darianovaxjo2012,darianovaxjo2012b}, at QCMC \cite{darianoQCMC2012b}, and other conferences. The general philosophy of the program have been object of four FQXi essays \cite{dariano2011essay,dariano2012essay,dariano2013essay,dariano2014essay}
partly republished in \cite{d2012essayADV,2012FQXi-springer,2013FQXi-springer,darianolaresearch}.}  Originally,  the idea of foliations over the quantum circuit has been explored, showing how the Lorentz time-dilations and space-contractions emerge by changing the foliation. This work has lead to the  analyses of Alessandro Tosini in Refs. \cite{d2010space,d2011emergence} and in the conclusive work \cite{d2013emergence}. However, it was soon realized that the foliation on the quantum circuit explores only the causal connectivity of the automaton, and works in the same way also for a classical circuit, as it happens for random walks in one dimension (see \eg Ref. \cite{stein1996concept}). Moreover, only rational boosts can be used, with the additional artifact that the events have to be coarse-grained in a boost-dependent way, with very different coarse-graining for very close values of the boost. This makes the recovery of the usual Lorentz transformations at large scales practically unfeasible. On the other hand, the first Dirac automaton in one space dimension \cite{mauro2012quantum} exhibited perfect Lorentz covariance for small wavevectors, which made clear that the quantum nature of the circuit plays a pivotal role in recovering the Lorentz invariance. In the same Ref. \cite{mauro2012quantum} it also emerged that the Dirac mass has to be upper bounded as a consequence of unitarity. 

The idea that so-called ``conventional''principles as homogeneity and isotropy may play a special role entered the scene since the very beginning \cite{darianopirsa-GUT1} through the connection with the old works of Ignatowsky \cite{von1910einige}, whereas ideas about how to treat gauge theories emerged already in Ref. \cite{darianopirsa-GUT3}. However, the project remained stuck for a couple of years because of two dead ends. First, we were looking to the realization of the quantum cellular automaton in terms of circuit gates, and we much later realized that the problem of connecting the gate realization (socalled Margolus scheme \cite{toffoli1987cellular}) to the linear quantum walk was a highly non trivial problem for dimension greater than one. Second, we where considering Jordan-Wigner mappings between local qubits and discrete Fermions \cite{darianosaggiatore}, generalizing to dimensions $d>1$ what can be done for d=1, and later Tosini realized that for $d>1$ such mapping cannot be done iso-locally \cite{IJMP14,EPL14}, namely preserving the locality of interactions. Paolo Perinotti, inspired from the work of Bialynicki-Birula \cite{bialynicki1994weyl}, recognized the first Dirac quantum walks in 2 and 3 dimensions. Later the graph structure of the walk was pointed out to be a Cayley graph of a group by Matt Brin, and the work of the derivation from principles of Weyl and Dirac \cite{PhysRevA.90.062106} followed after a Paolo's nontrivial solution of the unitarity conditions. This was the turning point of the whole program. It was soon recognized that the Maxwell field could be obtained by tensor product of two Weyl, and Alessandro Bisio soon found a way of achieving the photons with entangled pairs of Fermions. We finally realized the pivotal role played by the eigenvalue equation of the quantum in restating the relativity principle and recovering Lorentz covariance, and Bisio found the construction recovering the notion of particle as invariant of the deformed Poincar\'e group. 

\section{My encounter with David Ritz Finkelstein}
\label{sec:myencounter}
Vieque Island, January 6th 2014: FQXi IV International Conference on {\em The Physics of
  Information}.  The conference is very interactive, mostly devoted to debates, round tables, and
working groups. Max Tagmark organizes and chaires a morning session made of five-minutes talks.  The
audience includes distinguished scientists, a unique opportunity for presenting my Templeton project
{\em A Quantum-Digital Universe}. I want to say many things that I consider very important, and I
prepare my talk carefully, measuring the time of each single sentence, and memorizing each single
word. The result goes beyond my best expectations, with gratifying comments by a number of
scientists, some whom I meet for the first time.\footnote{Very flattering are the compliments of
  Federico Faggin, the designer of the first microprocessor at Intel.} But the best that happens is
that a beautiful old man, whom I never met before, with a white bear and a hat, literally
embraces me with a great smile, and almost with tears in his blue eyes says that I realized one of
his dreams. His enthusiasm, so passionate and authentic captures me. I spend most of the following
days discussing with him.  He invites me to visit him in his home in Atlanta.

I visit David on March 16th and 17th in a weekend during a visit in Boston. His house is beautiful,
with large windows opened on a surrounding forest.  With his wife Shlomit we have pleasant
conversations, some about their past encounter with the Dalai Lama.

\begin{figure}
 \includegraphics[width=1\textwidth]{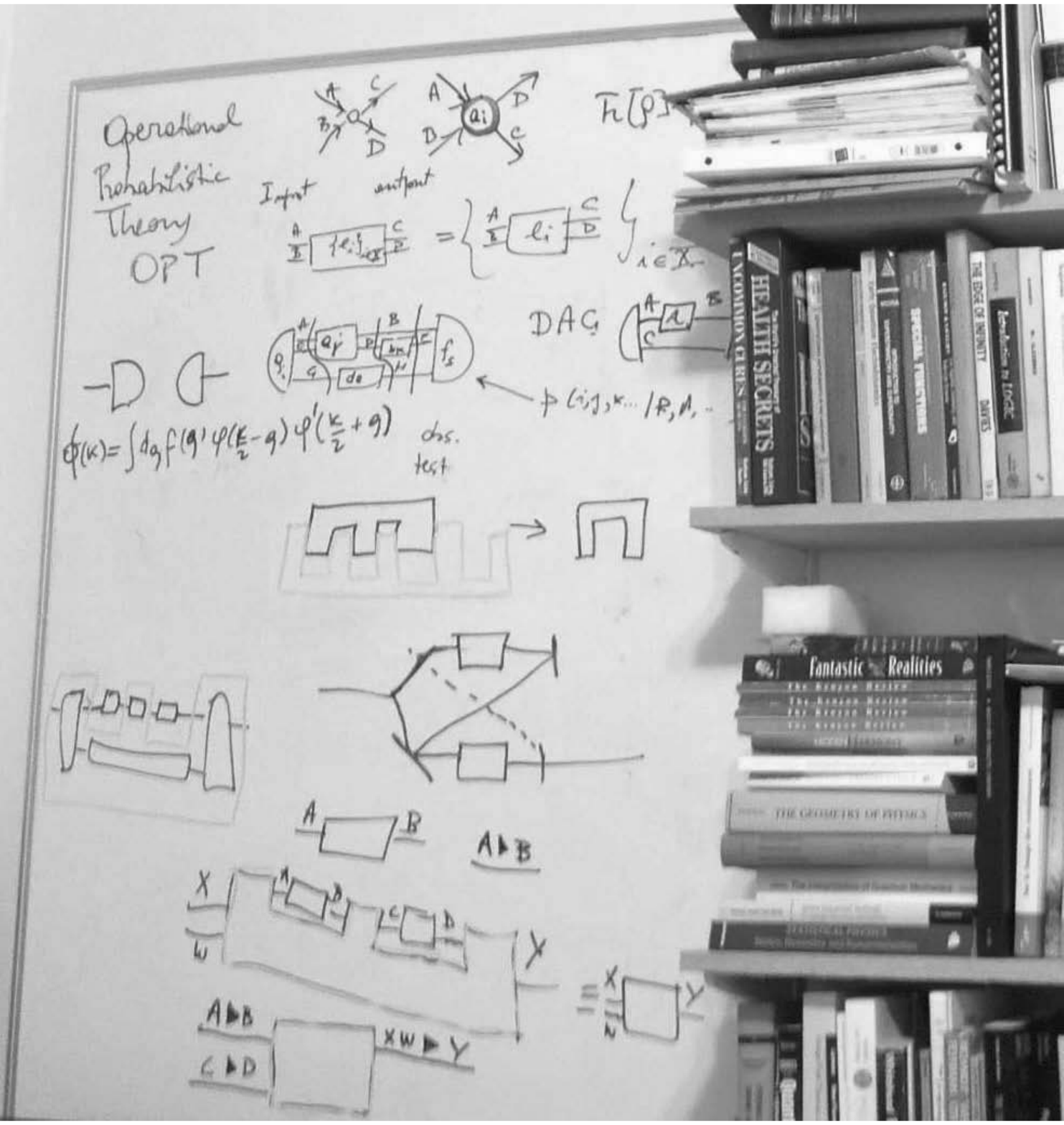}
\caption{The blackboard in David room, after a day-long tour on the derivation of quantum theory
  from principles \cite{QTfromprinciples}. You can notice some diagrams that pertain noncausal variations of quantum theory.\label{fig:blackboard}}
\end{figure}

David writes a nice dedication on my copy of his last book \cite{finkelstein2012quantumc}. He then asks me to explain to him the derivation of quantum theory from informa\-tion-theoretical
principles (which I did with my former students Paolo Perinotti and Giulio Chiribella
\cite{chiribella2011informational}: a textbook from Cambridge University Press is now in press
\cite{QTfromprinciples}).  I spend almost the two entire days in front of a small blackboard in
David room full of books (see Fig. \ref{fig:blackboard}), drawing diagrams and answering to his many
questions. His genuine interest will boost my enthusiasm for the years to come. 

After that visit David and I will continue to exchange emails. David regularly will send to me
updates of his work. We promise to exchange visits soon, but unfortunately this will not happen
again.
\section{My talk at FQXi 2014 {\em verbatim}}\label{sub:mytalk}
I'd like to tell you about the astonishing power of taking information more fundamental than matter,
the informational paradigm advocated by Wheeler, Feynman, and Seth Lloyd of ``the universe as a
huge quantum computer''. Quantum Theory is indeed a theory of information, since it can be
axiomatically derived from six axioms of pure information-theoretical nature. Five of the postulates
are in common with classical information. The one that discriminates between quantum and classical
is the principle of conservation of information--technically the purification postulate. Information
means describing everything in terms of input-output relations between events/transformations,
mathematically associating probabilities to closed circuits between preparations and observations.
[Some of the principles are conceptually quite new and interesting, such as the local
discriminability one, which in the quantum case reconciles holism with reductionism, with the
possibility of achieving complete information by local observation.]

Now these postulates provide only the quantum theory of abstract systems, not the mechanical part of
the theory. In order to get this you need to add new principles that lead to quantum field theory,
without assuming relativity and space-time. These principles describe the topology of interactions,
which determine the flow of information along the circuit. The first of these requirements is that
of finite info-density, corresponding to having a numerable set of finite-dimensional quantum
systems in interaction. Such principle, along with the assumption of unitarity, locality,
homogeneity, isotropy and minimal dimension of the systems in interaction, are equivalent to
minimizing the quantum algorithmic complexity of the information processing, reducing the physical
law to a bunch of few quantum gates, and leading to a description in terms of a Quantum Cellular
Automaton.

Now, it turns out that from these few assumptions only two quantum cellular automata follow that are connected by CPT, and
Lorentz covariance is broken. They both converge to the Dirac equation in the relativistic limit of
small masses and small wave-vectors. In the ultra-relativistic limit of large wave-vectors or masses
(corresponding to a Planck scale) Lorentz covariance becomes only an approximate symmetry, and one
has an energy scale and length scale that are invariant in addition to the speed of light,
corresponding to the Doubly Special Relativity of Amelino-Camelia/Smolin/Magueijo, with the
phenomenon of relative locality, namely that also coincidence in space, not only in time, is
observer-dependent. The covariance is given by the group of transformation leaving the dispersion
relations of the automaton invariant, and holds for energy-momentum.  When you get back to
space-time via Fourier, then you recover a space-time of quantum nature, with space-time points in
superposition.

The quantum cellular automaton can be regarded as a theory unifying scales ranging from Planck to Fermi. It is interesting
to notice that the same quantum cellular automaton also gives the Maxwell field, interestingly in the form of the de
Broglie-Fermi neutrino theory of the photon. With the principle of bounded information density, also
the Boson becomes an emergent notion, but the relation with Fermions is subtle in terms of
localization. The fact that the theory is discrete avoids all problems that plague quantum field
theory arising from the continuum, especially the problem of localization, but, most relevant, the
theory is quantum {\em ab initio}, with no need of quantization rules. And this is the great bonus of
taking information as more fundamental than matter.

\begin{acknowledgements}
  The present long-term unconventional project has needed a lot of energy and determination in the
  steps that had to be faced within the span of more than seven years. The work done up to now would
  have not been possible without the immeasurable contribution of some members of my group QUit in
  Pavia, as it can be seen from the list of references. All of them embraced with enthusiasm the
  difficult problems posed by the program, at the risk of their careers, in a authentically
  collaborative interaction. In particular, I am mostly grateful to Paolo Perinotti, with whom I had
  the most intense and interesting interactions of my entire career. I'm then very grateful to my
  postdocs Alessandro Bisio e Alessandro Tosini for their crucial extensive contribution, and to my
  PhD students Marco Erba and Nicola Mosco, and my previous PhD student Franco Manessi.  I am very
  grateful to my long-date friend Matt Brin for introducing me to some among the top mathematicians
  in geometric group theory, which otherwise it would have been impossible for me to meet. In
  particular: Benson Farb, Dennis Calegari, Cornelia Drutu, Romain Tessera, and Roberto Frigerio. I
  personally learnt a lot from Benson Farb in four meetings in at the Burgeois Pig caf\'e in
  Chicago, during my august visits at NWU in Evanston, and am grateful to Dennis Calegari for two
  interesting meetings at UC.  With Paolo Perinotti and Marco Erba we have visited Cornelia Drutu in
  Oxford, Romain Tessera in Paris, and Roberto Frigerio in Pisa, and from them we could learnt fast
  crucial mathematical notions and theorems, which otherwise it would have taken ages for us to find
  in books and articles.  

  I want then to acknowledge some friends that enthusiastically supported me
  in the difficult stages of the advancement of this program, in particular my mentor and friend
  Attilio Rigamonti, and my friends Giorgio Goggi, Catalina Curceanu, Marco Genovese, all of them
  experimentalists, along with the theoreticians Lee Smolin, Rafael Sorkin, Olaf Dryer, Lucien Hardy, Kalamara
  Fotini Markopoulou, Bob Coecke, Tony Short, Vladimir Buzek, Renato Renner, Wolfgang Schleich, Lev
  B. Levitin, and Andrei Khrennikov, for appreciating the value of this research since the
  earlier heuristic stage.  For inspiring scientific discussions I like to acknowledge Seth Lloyd,
  Reinhard F. Werner, Norman Margolus, Giovanni Amelino-Camelia, Shahn Majid, Louis H.  Kauffman,
  and Carlo Rovelli, whereas I wish to thank Arkady Plotnitsky and Gregg Jaeger for very exciting
  discussions about history and philosophy of physics.  

  I want finally to remark again the great help that I got from David Finkelstein, of whom I have been
  honored to be friend, and whose enthusiasm have literally boosted the second part of this project.

  Financially I acknowledge the support of the John Templeton foundation, whithout which the present
  project could had never take off from the preliminary heuristic stage.

\end{acknowledgements}


\bibliographystyle{spphys}
\bibliography{bibliography} 

\begin{thebibliography}{10}
\providecommand{\url}[1]{{#1}}
\providecommand{\urlprefix}{URL }
\expandafter\ifx\csname urlstyle\endcsname\relax
  \providecommand{\doi}[1]{DOI \discretionary{}{}{}#1}\else
  \providecommand{\doi}{DOI \discretionary{}{}{}\begingroup
  \urlstyle{rm}\Url}\fi

\bibitem{chiribella2011informational}
G.~Chiribella, G.M. D'Ariano, P.~Perinotti, Phys. Rev. A \textbf{84}, 012311
  (2011)

\bibitem{QTfromprinciples}
G.M. D'Ariano, G.~Chiribella, P.~Perinotti, \emph{Quantum Theory from First
  Principles} (Cambridge University Press, Cambridge, 2017).
\newblock In press

\bibitem{PhysRevA.90.062106}
G.M. D'Ariano, P.~Perinotti, Phys. Rev. A \textbf{90}, 062106 (2014)

\bibitem{bisio2013dirac}
A.~Bisio, G.M. D'Ariano, A.~Tosini, Phys. Rev. A \textbf{88}, 032301 (2013)

\bibitem{Bisio2015244}
A.~Bisio, G.M. D'Ariano, A.~Tosini, Annals of Physics \textbf{354}, 244  (2015)

\bibitem{Bisio2016}
A.~Bisio, G.M. D'Ariano, P.~Perinotti, Annals of Physics \textbf{368}, 177
  (2016)

\bibitem{susskind1995world}
L.~Susskind, Journal of Mathematical Physics \textbf{36}, 6377 (1995)

\bibitem{bousso2002holographic}
R.~Bousso, Reviews of Modern Physics \textbf{74}, 825 (2002)

\bibitem{john1955mathematical}
J.V. Neumann, \emph{Mathematical foundations of quantum mechanics}.
\newblock 2 (Princeton university press, 1955)

\bibitem{Giuliobook}
G.~Chiribella, G.M. D'Ariano, P.~Perinotti, in \emph{Quantum Theory:
  Informational Foundations and Foils}, ed. by G.~Chiribella, R.~Spekkens
  (Springer, 2016), pp. 165--175

\bibitem{shortreviewQT}
G.M. D'Ariano, P.~Perinotti, Found. Phys. \textbf{46}, 269 (2016)

\bibitem{halvorson2002no}
H.~Halvorson, R.~Clifton, Philosophy of Science \textbf{69}, 1 (2002)

\bibitem{kuhlmann2015real}
M.~Kuhlmann, Scientific American \textbf{24}, 84 (2015)

\bibitem{reviewderiv}
A.~Bisio, G.M. D'Ariano, P.~Perinotti, A.~Tosini, Found. Phys. \textbf{45},
  1137 (2015)

\bibitem{reviewanaly}
A.~Bisio, G.M. D'Ariano, P.~Perinotti, A.~Tosini, Found. Phys. \textbf{45},
  1203 (2015)

\bibitem{mauro2012quantum}
G.M. D'Ariano, Phys. Lett. A \textbf{376}(5), 697 (2012)

\bibitem{lrntz3d}
A.~Bisio, G.M. D'Ariano, P.~Perinotti, arXiv preprint arXiv:1503.01017v3
  (2015)

\bibitem{FOPDP16}
G.M. D'Ariano, P.~Perinotti, Front. Phys. \textbf{12}, 120301 (2017).
\newblock \doi{10.1007/s11467-016-0616-z}.
\newblock Special Topic: Quantum Communication, Measurement, and Computing, in
  press (arXiv:1608.02004)

\bibitem{d2015virtually}
G.M. D'Ariano, M.~Erba, P.~Perinotti, A.~Tosini, arXiv preprint
  arXiv:1511.03992  (2015)

\bibitem{hey1998feynman}
A.J. Hey, \emph{Feynman and Computation-Exploring the Limits of Computers}
  (Perseus Books, 1998)

\bibitem{Bravyi2002210}
S.B. Bravyi, A.Y. Kitaev, Annals of Physics \textbf{298}, 210  (2002)

\bibitem{IJMP14}
G.M. D'Ariano, F.~Manessi, P.~Perinotti, A.~Tosini, Int. J. Mod. Phys. A
  \textbf{17}, 1430025 (2014)

\bibitem{EPL14}
G.M. D'Ariano, F.~Manessi, P.~Perinotti, A.~Tosini, Europhysics Letters
  \textbf{107}, 20009 (2014)

\bibitem{colodny1986quarks}
R.G. Colodny (ed.), \emph{From quarks to quasars: philosophical problems of
  modern physics} (University of Pittsburgh Pre, 1986)

\bibitem{harpe}
P.~de~La~Harpe, \emph{Topics in geometric group theory} (University of Chicago
  Press, 2000)

\bibitem{Farb}
B.~Farb, private communication

\bibitem{Drutu}
C.~Drutu, private communication

\bibitem{Tessera}
R.~Tessera, private communication

\bibitem{unpubDEP}
G.M. D'Ariano, M.~Erba, P.~Perinotti, unpublished

\bibitem{Kapovich-notes}
M.~Kapovich, unpublished

\bibitem{mackey1951induced}
G.W. Mackey, American Journal of Mathematics \textbf{73}, 576 (1951)

\bibitem{mackey1952induced}
G.W. Mackey, Annals of Mathematics \textbf{55}, 101 (1952)

\bibitem{mackey1953induced}
G.W. Mackey, Annals of Mathematics \textbf{58}, 193 (1953)

\bibitem{d2016discrete}
G.M. D'Ariano, N.~Mosco, P.~Perinotti, A.~Tosini, Entropy \textbf{18}, 228
  (2016)

\bibitem{path1d}
G.M. D'Ariano, N.~Mosco, P.~Perinotti, A.~Tosini, Phys. Lett. A \textbf{378}
  (2014)

\bibitem{path2d}
G.M. D'Ariano, N.~Mosco, P.~Perinotti, A.~Tosini, EPL \textbf{109} (2015)

\bibitem{path3d}
G.M. D'Ariano, N.~Mosco, P.~Perinotti, A.~Tosini,  (2015).
\newblock Discrete Feynman propagator for the Weyl quantum walk in 3+1
  dimensions

\bibitem{kauffman1996discrete}
L.H. Kauffman, H.P. Noyes, Physics Letters A \textbf{218}(3), 139 (1996)

\bibitem{greenberger1970theory}
D.M. Greenberger, Journal of Mathematical Physics \textbf{11}(8), 2329 (1970)

\bibitem{ellis1992string}
J.~Ellis, N.~Mavromatos, D.V. Nanopoulos, Physics Letters B \textbf{293}(1), 37
  (1992)

\bibitem{lukierski1995classical}
J.~Lukierski, H.~Ruegg, W.J. Zakrzewski, Annals of Physics \textbf{243}(1), 90
  (1995)

\bibitem{Quantidischooft1996}
G.~'t~Hooft, Class. Quantum Grav. \textbf{13}(5), 1023 (1996)

\bibitem{amelino2001testable}
G.~Amelino-Camelia, Physics Letters B \textbf{510}(1), 255 (2001)

\bibitem{PhysRevLett.88.190403}
J.~Magueijo, L.~Smolin, Phys. Rev. Lett. \textbf{88}, 190403 (2002)

\bibitem{amelino1998tests}
G.~Amelino-Camelia, J.~Ellis, N.~Mavromatos, D.V. Nanopoulos, S.~Sarkar, Nature
  \textbf{393}(6687), 763 (1998)

\bibitem{abdo2009limit}
A.~Abdo, M.~Ackermann, M.~Ajello, K.~Asano, W.~Atwood, M.~Axelsson, L.~Baldini,
  J.~Ballet, G.~Barbiellini, M.~Baring, et~al., Nature \textbf{462}(7271), 331
  (2009)

\bibitem{vasileiou2013constraints}
V.~Vasileiou, A.~Jacholkowska, F.~Piron, J.~Bolmont, C.~Couturier, J.~Granot,
  F.W. Stecker, J.~Cohen-Tanugi, F.~Longo, Physical Review D \textbf{87}(12),
  122001 (2013)

\bibitem{amelino2009prospects}
G.~Amelino-Camelia, L.~Smolin, Physical Review D \textbf{80}(8), 084017 (2009)

\bibitem{bibeau2013doubly}
A.~Bibeau-Delisle, A.~Bisio, G.M. D'Ariano, P.~Perinotti, A.~Tosini, EPL
  (2015).
\newblock In press

\bibitem{hopf}
A.~Bisio, G.M. D'Ariano, P.~Perinotti, Phil. Trans. R. Soc. A \textbf{374}
  (2016)

\bibitem{PhysRevD.16.3031}
L.~Susskind, Phys. Rev. D \textbf{16}, 3031 (1977)

\bibitem{amelino2001planck}
G.~Amelino-Camelia, T.~Piran, Physical Review D \textbf{64}(3), 036005 (2001)

\bibitem{amelino2002relativity}
G.~Amelino-Camelia, International Journal of Modern Physics D \textbf{11}(01),
  35 (2002)

\bibitem{magueijo2002lorentz}
J.~Magueijo, L.~Smolin, Physical Review Letters \textbf{88}(19), 190403 (2002)

\bibitem{Hubbard}
 (2012)

\bibitem{darianopirsa-GUT1}
G.M. D'Ariano, A computational grand-unified theory (2010).
\newblock URL:http://pirsa.org/10020037

\bibitem{darianopirsa-GUT2}
G.M. D'Ariano, Physics as information: Quantum theory meets relativity (2010).
\newblock Http://pirsa.org/10110080

\bibitem{darianopirsa-GUT3}
G.M. D'Ariano, A quantum-digital universe (2011).
\newblock Http://pirsa.org/11050042

\bibitem{darianovaxjo2010}
G.M. D'Ariano, in \emph{Quantum Theory: Reconsideration of Foundations 5}, vol.
  CP1232 (AIP, 2010), vol. CP1232, p.~3

\bibitem{darianovaxjo2011}
G.M. D'Ariano, in \emph{Advances in Quantum Theory}, vol. CP1327 (AIP, 2011),
  vol. CP1327, p.~7

\bibitem{darianovaxjo2012}
G.M. D'Ariano, in \emph{Foundations of Probability and Physics - 6}, vol.
  CP1508 (AIP, 2012), vol. CP1508, p. 146

\bibitem{darianovaxjo2012b}
G.M. D'Ariano, in \emph{Foundations of Probability and Physics - 6}, vol.
  CP1424 (AIP, 2012), vol. CP1424, p. 371

\bibitem{darianoQCMC2012b}
G.M. D'Ariano, in \emph{Quantum Communication, measurement and Computing
  (QCMC)}, vol. CP1363 (AIP, 2011), vol. CP1363, p.~63

\bibitem{dariano2011essay}
G.~D'Ariano, FQXi Essay Contest: {\em Is Reality Digital or Analog?}  (2011)

\bibitem{dariano2012essay}
G.~D'Ariano, FQXi Essay Contest: {\em Questioning the Foundations: Which of Our
  Basic Physical Assumptions Are Wrong?}  (2012)

\bibitem{dariano2013essay}
G.~D'Ariano, FQXi Essay Contest: {\em It From Bit or Bit From It?}  (2013)

\bibitem{dariano2014essay}
G.~D'Ariano, FQXi Essay Contest: {\em Trick or Truth: the Mysterious Connection
  Between Physics and Mathematics}  (2013)

\bibitem{d2012essayADV}
G.M. D'Ariano, Adv. Sci. Lett. \textbf{17}, 130 (2012)

\bibitem{2012FQXi-springer}
G.M. D'Ariano, in \emph{Questioning the Foundations of Physics, The Frontiers
  Collection}, ed. by A.A. et~al. (Springer, 2015), pp. 165--175

\bibitem{2013FQXi-springer}
G.M. D'Ariano, in \emph{It From Bit or Bit From It?, The Frontiers Collection},
  ed. by A.A. et~al. (Springer, 2015), pp. 25--35

\bibitem{darianolaresearch}
G.M. D'Ariano, La Recherche \textbf{489}, 48 (2014)

\bibitem{d2010space}
G.M. D'Ariano, A.~Tosini, arXiv preprint arXiv:1008.4805  (2010)

\bibitem{d2011emergence}
G.M. D'Ariano, A.~Tosini, arXiv preprint arXiv:1109.0118  (2011)

\bibitem{d2013emergence}
G.M. D'Ariano, A.~Tosini, Studies in History and Philosophy of Science Part B:
  Studies in History and Philosophy of Modern Physics \textbf{44}(3), 294
  (2013)

\bibitem{stein1996concept}
I.~Stein, \emph{The concept of object as the foundation of physics} (Peter
  Lang, 1996)

\bibitem{von1910einige}
W.~von Ignatowsky, Verh. Deutsch. Phys. Ges \textbf{12}, 788 (1910)

\bibitem{toffoli1987cellular}
T.~Toffoli, N.~Margolus, \emph{Cellular automata machines} (MIT press, 1987)

\bibitem{darianosaggiatore}
G.M. D'Ariano, Il Nuovo Saggiatore \textbf{28}, 13 (2012)

\bibitem{bialynicki1994weyl}
I.~Bialynicki-Birula, Physical Review D \textbf{49}(12), 6920 (1994)

\bibitem{finkelstein2012quantumc}
D.R. Finkelstein, \emph{Quantum relativity: a synthesis of the ideas of
  Einstein and Heisenberg} (Springer Science \& Business Media, 2012)

\end{thebibliography}

\end{document}